\pdfoutput=1
\documentclass[11pt,onecolumn,superscriptaddress,nofootinbib,floatfix]{revtex4-1}
\usepackage{graphicx}
\usepackage{epsfig}
\usepackage{amsmath}
\usepackage{amsfonts}
\usepackage{float}
\usepackage{amssymb}
\usepackage{color}
\usepackage[bottom]{footmisc}
\usepackage{array,multirow,makecell}
\usepackage{slashed}
\usepackage[colorlinks,citecolor=blue]{hyperref}
\usepackage{ulem}
\usepackage[utf8]{inputenc}
\usepackage[english]{babel}
\usepackage{url}
\usepackage{hyperref}
\usepackage{xcolor}
\usepackage{subfigure}
\usepackage{lipsum}
\usepackage[bottom]{footmisc}
\usepackage{pifont}
\def\lsim{\raise0.3ex\hbox{$\;<$\kern-0.75em\raise-1.1ex\hbox{$\sim\;$}}}
\def\gsim{\raise0.3ex\hbox{$\;>$\kern-0.75em\raise-1.1ex\hbox{$\sim\;$}}}

\def\bea{\begin{eqnarray}}
\def\eea{\end{eqnarray}}

\DeclareUnicodeCharacter{2212}{-}
\begin{document}
\title{Discriminating and Constraining the Synchrotron and Inverse Compton Radiations from Primordial Black Hole and Dark Matter at the Galactic Centre Region}

\author{Upala Mukhopadhyay}
\email[E-mail: ]{upala.mukhopadhyay@saha.ac.in}
\affiliation{Theory Division, Saha Institute of Nuclear Physics, 1/AF Bidhannagar, Kolkata 700064, India.}
\affiliation{Homi Bhabha National Institute, Training school complex,  Anushaktinagar,
Mumbai 400094, India.}

\author{Debasish Majumdar}
\email[E-mail: ]{debasish.majumdar@saha.ac.in}
\affiliation{Theory Division, Saha Institute of Nuclear Physics, 1/AF Bidhannagar, Kolkata 700064, India.}
\affiliation{Homi Bhabha National Institute, Training school complex,  Anushaktinagar,
Mumbai 400094, India.}

\author{Avik Paul}
\email[E-mail: ]{avik.paul@saha.ac.in}
\affiliation{Theory Division, Saha Institute of Nuclear Physics, 1/AF Bidhannagar, Kolkata 700064, India.}
\affiliation{Homi Bhabha National Institute, Training school complex,  Anushaktinagar,
Mumbai 400094, India.}
\begin{abstract}
The evaporations of Primordial Black Holes (PBH) (via Hawking radiation) can produce electrons/positrons ($e^-/e^+$) in the Galactic Centre (GC) region which under the influence of the magnetic field of Centre region can emit synchrotron radiation. These $e^-/e^+$ can also induce Inverse Compton radiation due to the scattering with ambient photons. In this work three different PBH mass distributions namely, monochromatic, power law and lognormal distributions are considered to calculate such radiation fluxes. On the other hand, annihilation or decay of dark matter in the Galactic Centre region can also yield $e^-/e^+$ as the end product which again may emit synchrotron radiation in the Galactic magnetic field and also induce Inverse Compton scattering. In this work a comparative study is made for these radiation fluxes from both PBH evaporations and from dark matter origins and their detectabilities are addressed in various ongoing and other telescopes as well as in upcoming telescopes such as SKA. Moreover, constraints on the model parameters are obtained from these experimental predictions. The variations of these radiation fluxes with the distance from the Galactic Centre are also computed and it is found that such variations could be a useful probe to determine the mass of PBH or the mass of dark matter.
\end{abstract}
\maketitle
\section{Introduction} \label{intro}
Primordial Black Holes (PBHs) \cite {10.1093/mnras/152.1.75, 10.1093/mnras/168.2.399} are massive compact halo objects that could have been created shortly after Big Bang. Their masses depend on the time at which they are created and therefore they are likely to have a range of masses or a mass distribution instead of a single mass for all the PBHs. The PBHs being nearly collisionless and could be stable if sufficiently 
massive can well qualify to be candidates for dark matter \cite{PhysRevD.94.083504, Belotsky_2014, Khlopov_2010, PhysRevD.81.104019}. In fact it is suggested from observational hints \cite{Clesse:2017bsw}, found from the abundance studies of PBHs,
that the PBHs may constitute the dark matter content or a component of the dark matter content. In fact several observational results such as Voyager 1 data \cite{Boudaud:2021zzd, Cummings_2016}, strong lensing \cite{Katz_2018, Montero_Camacho_2019} 
measurements, background gamma ray data \cite{Carr:2016hva, Laha:2019ssq} etc. indicate a fractional mass ratio $f_{\rm PBH}$ of the total PBHs to the total dark matter \cite{Carr:2016hva, Laha:2019ssq, Carr:2020gox, Belotsky_2019, Boudaud:2018hqb}. This indicates that PBHs may constitute a fraction of total dark matter content and that may vary with the PBH masses. PBHs are believed to undergo evaporation via Hawking radiation. If PBH mass is sufficiently small ($\lesssim 10^{14}$ g), they would be withered away completely via Hawking radiation. The Standard Model particles such as electrons/positrons may be emitted via Hawking radiation and these may interact with surrounding medium. These electrons/positrons could produce synchrotron radiation by the influence of the magnetic field present in the medium. These particles may also induce Inverse Compton scattering with the ambient photons and these IC scattered photons also carry information regarding the nature and evaporation of PBHs.

On the other hand the indirect detection of dark matter \cite{Fermi-LAT:2016uux, Cavasonza_2017, Storm_2013, Huang_2020, Egorov_2013} is related to detecting the signal from the products of the dark matter annihilation or decay. The gravity of massive astrophysical objects may trap dark matter from escaping those objects and when they are trapped in substantial magnitude inside the core of the massive objects such as Galactic Centre, Solar core etc. they can undergo the process of self annihilation to produce standard model particles such as fermions, photons etc. as the end products. The excess signal of these fermions, photons etc.
could be a probable signature of dark matter indirect detections. Possible decay of dark matter can also lead to fermions as one of the final products. 
If electrons/positrons are the products of such dark matter annihilations or dark matter decays then these charged particles under the influence of Galactic magnetic field can also emit synchrotron radiation which may be detected by terrestrial radio telescopes. On the other hand these electrons/positrons can also induce Inverse Compton (IC) scattering with ambient photons and the IC scattered photons may again be a useful probe for these phenomena.

Thus synchrotron radiations from distant cosmos could be an effective probe 
to understand unknown phenomena in the Universe. The present and future radio telescopes such as Square Kilometer Array (SKA) \cite{SKA2017, Kar:2019cqo, Bertolami:2018lel, Kar:2018rlm}, GMRT \cite{uGMRT2017}, uGMRT \cite{uGMRT2017}, MeerKAT \cite{Brederode:2018gh}, LOFAR \cite{van_Haarlem_2013}, Jodrell Bank \cite{10.1093/mnras/177.2.319}, JVLA \cite{JVLA}, ASKAP \cite{SKA2017} etc. provide new possibilities of probing the evidence and physics of dark matter and PBHs among other cosmological phenomena.

In the present work we address these issues of emission of synchrotron and IC radiations from the products of PBH evaporation in our Galaxy as well as the synchrotron and IC emissions resulting from the end products of possible dark matter self annihilation or dark matter decay. The synchrotron radiation fluxes are computed for PBH decays and for both dark matter cases (annihilation and evaporation as well as dark matter decay). The IC radiation flux for these cases are also calculated.
The results are then compared. Various observational results of existing radio telescopes and the sensitivity estimates given by SKA for 100 hour and 1000 hour runs are then addressed to asses the detectabilities of the computed signals.
 
For both the cases (synchrotron and IC), two possibilities are considered for electrons/positrons - high energy electrons/positrons (of the order of GeV) and low energy electrons/positrons (of the order of MeV). For the case of dark matter, two dark matter masses namely 100 GeV and 50 GeV are considered for demonstrative purpose and the dark matter here is considered to WIMPs or Weakly Interacting Massive Particles. For the PBH case, three categories of PBHs are considered. The first category includes constant PBH masses ($M_{\rm PBH}$) and the constant masses are chosen to be the two extremum masses, namely $10^{14}$ g and $10^{17}$ g, of the range of PBH masses considered in this work. For the other two categories, we consider mass distributions for PBHs. In the second category, a power law mass distribution \cite{Carr:2017jsz, Chan:2020zry} has been adopted while we analyse with  a lognormal distribution of PBH masses \cite{Carr:2017jsz, Chan:2020zry} in the third category. In Ref. \cite{Chan:2020zry} the authors have addressed the synchrotron emissions originated from the PBH evaporation at the Galactic Centre and have constrained the PBH to dark matter ratio in case of three above mentioned PBH mass distributions using the radio observational data of the inner Galactic Centre.
In this work, we compute not only the synchrotron emission but also the IC radiation from evaporation of PBHs (for the three possible PBH mass distributions mentioned above) at the Galactic Centre region. We also compute the flux densities of synchrotron and IC radiations originated from the decay/annihilation of dark matter. In addition, the parameters of PBH mass distributions as well as dark matter mass and decay/annihilation rates are constrained in this work from the observational data namely SKA, GMRT, uGMRT, Jodrell Bank, MeerKat, VLA,
JVLA, LOFAR, ASAP.
Moreover, for all these cases, the variations of synchrotron flux (and IC radiations) with the distance from the Galactic Centre are also estimated.

The paper is organised as follows. In Sect. \ref{formalism}, we provide detailed theoretical framework for computations of synchrotron radiations and IC radiations for the cases of dark matter and PBHs as described above. Sect. \ref{formalism} also furnishes the analytical expressions  for the estimations magnetic 
field in and around the Galactic Centre. Sect. \ref{formalism} therefore consists of 
various subsections. In Sect. \ref{result} and in the subsections within, we furnish 
detailed calculational procedure and present our results. Finally in Sect. \ref{discussion}, we conclude with a summary.
 \vspace{-10.04mm}
 \vspace{1cm}
\section{Calculations for Synchrotron Radiation and Inverse Compton Radiation}\label{formalism}
In this section, we estimate synchrotron flux and Inverse Compton (IC) flux originated due to possible decay of primordial black holes (PBH) or the decay/ annihilation of dark matter (DM) at the Galactic Centre. The possible decay of PBH or the possible decay/annihilation of DM in the Galactic Centre may produce electrons ($e^-$) and positrons ($e^+$) as the final products. In the presence of large magnetic field in the Galactic Centre region these $e^-/ e^+$ produce synchrotron radiations while the scattering of $e^-/ e^+$ with surrounding photons gives rise to IC radiations.

\subsection{Diffusion Equation}
In order to obtain synchrotron flux and IC flux as mentioned above, we need to solve the diffusion equation with the form given by \cite{Hooper:2012jc},
\begin{equation}
K(E)\nabla^2 \frac{d n_e(E,r)}{dE} +\frac{\partial}{\partial E}\left[b(E,r) \frac{d n_e (E,r)}{d E}\right] +Q(E,r)=0\,\,. \label{diffusion}
\end{equation}
Here, $\frac{d n_e(E,r)}{dE}$ is the number density of electron (positron) per unit energy interval at position $r$, $K(E)$ and $b(E,r)$ represent diffusion coefficient and energy loss coefficient respectively while $Q(E,r)$ denotes the electron (positron) source term. We assume the system is in equilibrium and a steady state solution of the above equation is being sought. To this end we consider $\frac{d n_e(E,r)}{dE}$ is independent of time.

The produced $e^-$ and $e^+$ can lose their energy via different processes such as synchrotron, IC, bremsstrahlung and Coulomb. Energy loss depends on the magnetic field strength (synchrotron loss), CMB photon spectrum (IC loss), the electron density $n_e$ and hydrogen density $n_H$ (bremsstrahlung and Coulomb losses) as well the energies of $e^-$ and $e^+$. Total energy loss term $b(E,r)$ can be expressed as \cite{Paul:2018njd}, 
\begin{equation}
b(E,r)=b_{\rm synchrotron}(E,r)+b_{\rm IC}(E,r)+b_{\rm bremsstrahlung}(E,r) + b_{\rm coulomb}(E,r)\,\,,
\label{loss}
\end{equation}
with 
\cite{Kar:2019cqo,Linden:2010eu,Laha:2012fg,Colafrancesco:2015ola,Springel:2008cc}
\begin{equation}
b_{\rm synchrotron}(E,r) = \frac{dE}{dt}\vert_{\rm synchrotron} = 3.4 \times 10^{-17} {\rm GeVs^{-1}} \left(\frac{E}{{\rm GeV}}\right)^2 \left(\frac{B}{3{\rm \mu G}}\right)^2\,\,,
\end{equation}

\begin{equation}
b_{\rm IC}(E,r) = \frac{dE}{dt}\vert_{\rm IC} = 10^{-16} {\rm GeVs^{-1}} \left(\frac{E}{{\rm GeV}}\right)^2 \left(\frac{U_{rad}}{{\rm eV cm^{-3}}}\right)\,\,,
\end{equation}

\begin{equation}
b_{\rm bremsstrahlung}(E,r) = \frac{dE}{dt}\vert_{\rm bremsstrahlung} = 3 \times 10^{-15} {\rm GeVs^{-1}} \left(\frac{E}{{\rm GeV}}\right) \left(\frac{n_H}{{3\rm cm^{-3}}}\right)\,\,,
\end{equation}

\begin{equation}
b_{\rm coulomb}(E,r) = \frac{dE}{dt}\vert_{\rm coulomb} = 6.13 \times 10^{-16} {\rm GeVs^{-1}} n_e \left[1+\frac{{\rm log}\frac{E/m_e}{n_e}}{75}\right]\,\,.
\end{equation}

In the above equations $m_e$ is the electron mass, $n_H$ denotes the number density of the hydrogen nuclei in the Milky Way while $U_{rad}$ is the radiation energy density.

Since magnetic field $B(r)$ in the neighbourhood of the Galactic Centre is very large, the contribution of the second term in Eq. \ref{diffusion} is much more significant than the contribution of the first term in the same equation. Thus generally the first term of Eq. \ref{diffusion} is neglected and the electron spectrum is obtained as \cite{Paul:2018njd},
\begin{equation}
\frac{d n_e}{dE} = \frac{\int Q(E^\prime, r) dE^\prime}{b(E,r)}\,\,.
\label{electron_spectrum}
\end{equation}

\subsection{Synchrotron Radiation}
As mentioned earlier, the electrons and positrons, produced from the decay of PBH or the decay/annihilation of DM, can generate synchrotron radiation at radio frequencies under the influence of the strong magnetic field $B$ 
at the Galactic Centre.

The synchrotron power density per unit frequency can be expressed as \cite{McDaniel:2017ppt},
\begin{equation}
j_{\rm sync}(\nu, r)=2 \int dE \frac{d n_e}{dE} P_{\rm sync}(\nu, E, r)\,\,, \label{jsync}
\end{equation}
where the synchrotron power $P_{\rm sync}(\nu, E, r)$ is defined as \cite{Longair2011},
\begin{equation}
P_{\rm sync}(\nu, E, r)=\frac{1}{4 \pi \epsilon_0} \frac{\sqrt{3}e^3 B(r)}{m_e c} f\left(\frac{\nu}{\nu_c}\right)\,\,.\label{psync}
\end{equation}
Here, $\nu_c$ is the critical frequency which is expressed by \cite{Bertone:2008xr},
\begin{equation}
\nu_c=\frac{3 e E^2 B}{4 \pi m_e^3 c^4}\,\,,
\end{equation} 
and the function $f(x)$ is defined as \cite{Longair2011},
\begin{equation}
f(x)=x \int_x^\infty dx^\prime K_{\frac{5}{3}}(x^\prime)\,\,, 
\end{equation}
with $K_{\frac{5}{3}}(x^\prime)$ is the modified Bessel's function of order $\frac{5}{3}$.

The synchrotron flux density can be obtained by taking the line of sight (l.o.s) integral of the synchrotron power density $j_{\rm sync}(\nu, r)$ and then subsequently taking integral over the solid angle ($\delta \Omega$) of the observed range. The average synchrotron flux density over the solid angle ($\delta \Omega$) is written as \cite{McDaniel:2017ppt,Cirelli:2010xx}
\begin{equation}
F_{\rm sync}=\frac{1}{4 \pi}\int d\Omega \int_{\rm l.o.s} dl j_{\rm sync}(\nu, r)\,\,.
\label{synflux}
\end{equation}
The integral over the line of sight $l$ and the solid angle $\delta \Omega$ can be calculated from the relations $r=(r_\odot^2 +l^2 -2 r_\odot l {\rm cos}\theta)^\frac{1}{2}$ and $\delta \Omega=2\pi \int_{\theta_{\rm min}}^{\theta_{\rm max}} d\theta {\rm sin}\theta$ respectively. Here, $r_\odot$ is the distance between the Sun and the Galactic Centre while $r$ represents the distance of the site of decay/annihilation from the Galactic Centre and $\theta$ denotes the angle between the direction of the l.o.s and the line joining the Earth and the Galactic Centre.

\subsection{Inverse Compton Radiation}
The electrons and positrons originated from the decay of PBHs or the decay/annihilation of DM can produce radiations from the process of Inverse Compton scattering with the background photons, dominantly with Cosmic Microwave Background photons with temperature 2.73 K. The low energy photons gain energy through IC scattering from the kinetic energy of $e^-$ or $e^+$.

The IC power can be written as \cite{McDaniel:2017ppt},
\begin{equation}
P_{\rm IC}(E, E_\gamma)= c E_\gamma \int d\epsilon n(\epsilon) \sigma(E,E_\gamma,\epsilon)\,\,,
\end{equation}
where $ \sigma(E,E_\gamma,\epsilon)$ and $n(\epsilon)$ respectively denote IC scattering cross section and number density of photon. In the above equation, $\epsilon$, $E$ and $E_\gamma$ represent energy of the target photons, energy of electrons or positrons and energy of the up-scattered photons respectively. 

IC scattering cross section $\sigma(E,E_\gamma,\epsilon)$ is calculated from the Klein-Nishina formula \cite{Klein1929}
\begin{equation}
\sigma(E,E_\gamma,\epsilon)=\frac{3 m_e^2 c^4 \sigma_T}{4\epsilon E^2} G(q,\Gamma^\prime)\,\,,
\end{equation}
where $\sigma_T$ is the Thompson scattering cross section and $G(q, \Gamma^\prime)$ is defined as \cite{Blumenthal:1970gc},
\begin{equation}
G(q, \Gamma^\prime)=\left[ 2q {\rm ln}q + (1+2q)(1-q) + \frac{(2q)^2 (1-q)}{2 (1+ \Gamma^\prime q)}\right]\,\,,
\end{equation}
with $\Gamma^\prime=\frac{4 \epsilon E}{m_e^2 c^4}$ and $q=\frac{E_\gamma}{\Gamma^\prime (E-E_\gamma)}$ .

Now, the IC power density per unit frequency can be calculated as \cite{Cirelli:2010xx},
\begin{equation}
j_{\rm IC}(E_\gamma, r)=2 \int dE \frac{d n_e}{dE} P_{\rm IC}(E, E_\gamma)\,\,,
\end{equation}
and the average IC flux density over the solid angle is derived as \cite{Cirelli:2010xx},
\begin{equation}
F_{\rm IC}=\frac{1}{4 \pi}\int d\Omega \int_{\rm l.o.s} dl j_{\rm IC}(E_\gamma, r)\,\,.
\label{ICflux}
\end{equation}
\subsection{Source Term of Electrons and Positrons Spectra}
In order to calculate synchrotron flux density or IC flux density from Eq. \ref{synflux} and Eq. \ref{ICflux}, one needs to obtain $\frac{dn_e(E,r)}{dE}$ (number density of electrons (positrons) per unit energy interval at position $r$) by solving Eq. \ref{electron_spectrum}. This can be done by defining $e^-/ e^+$ source term $Q(E,r)$ for the decay of PBH or the decay/annihilation of DM. For the decay and annihilation of DM, $Q(E,r)$ are defined as \cite{Dutta:2020lqc,Colafrancesco:2005ji},
\begin{equation}
Q_{\rm decay}(E,r)=\frac{\Gamma_{\rm DM}}{m_{\rm DM}}\rho(r) \frac{dN_e}{dE}\,\,,
\label{Qdecay}
\end{equation}
\begin{equation}
Q_{\rm annihilation}(E,r)=\frac{\left\langle   \sigma v \right\rangle  }{2 m_{\rm DM}^2}\rho^2(r) \frac{dN_e}{dE}\,\,.
\label{Qanni}
\end{equation}
Here, $\Gamma_{\rm DM}$ is the decay rate of DM and $\left\langle   \sigma v \right\rangle $ is the annihilation rate of DM while DM mass is denoted by $m_{\rm DM}$. In the above equations $\frac{dN_e}{dE}$ represents $e^-/ e^+$ energy distribution originated from the decay or annihilation of each DM.

For the decay of PBH the source term $Q(E,r)$ is written as \cite{Carr:2009jm}, \cite{Chan:2020zry}, \cite{Halzen:1995hu},
\begin{equation}
Q_{\rm PBH}(E,r)=\frac{f_{\rm PBH}}{M_{\rm PBH}}\rho(r) \frac{d\dot{N}_e}{dE}\,
\label{QPBH}
\end{equation}
where $\frac{d\dot{N}_e}{dE}$, the $e^-/ e^+$ energy distribution produced per unit time from the decay of a single PBH is given as,
\begin{equation}
\frac{d\dot{N}_e}{dE}=\frac{1}{2\pi \hbar}\frac{\Gamma_{\rm PBH}}{{\rm exp}(E/k_B T_{\rm PBH})+1}\,\,.
\end{equation} 
Here, $\Gamma_{\rm PBH}$ is the absorption coefficient of electron like particles (spin $\frac{1}{2}$ particles) and approximately expressed as \cite{MacGibbon:1990zk} $\Gamma_{\rm PBH}\simeq\frac{27 G^2 M_{\rm PBH}^2 E^2}{\hbar^2 c^6}\simeq 3.82 \times 10^{-2} \left(\frac{M_{\rm PBH}}{10^{13} {\rm g}}\right)^2 \left(\frac{E}{{\rm GeV}}\right)^2$. Temperature of a PBH ($T_{\rm PBH}$) is linked to its mass ($M_{\rm PBH}$) through the relation \cite{Hawking:1974rv} $k_B T_{\rm PBH} \simeq 1.06 \left(\frac{10^{13} {\rm g}}{M_{\rm PBH}}\right)$ GeV. In Eq. \ref{QPBH}, $f_{\rm PBH}$ represents the fraction of the PBH density to the total DM density $\rho(r)$ (i.e., $\rho_{\rm PBH}(r)=f_{\rm PBH} \rho(r)$).

In order to compute the $Q_{\rm PBH}(E,r)$ of Eq. \ref{QPBH} it is assumed that all the PBHs have equal mass value. But recent studies suggest that realistic production mechanisms could be due to an extended mass distribution of the PBHs. In this scenario $Q_{\rm PBH}(E,r)$ can be written as \cite{Boudaud:2018hqb}, \cite{Carr:2017jsz},
\begin{equation}
Q_{\rm PBH}(E,r)=\frac{f_{\rm PBH} \rho(r)}{\rho_\odot} \int_{\forall M_{\rm PBH}} d M_{\rm PBH} \frac{g(M_{\rm PBH})}{M_{\rm PBH}} \frac{d\dot{N}_e}{dE}\,\,,
\end{equation}
with $g(M_{\rm PBH})$ denotes mass distribution of PBHs normalised to $\rho_\odot$. In this work, we have considered two types of extended mass distribution of PBHs namely, power law distribution and lognormal distribution.

The power law mass distribution of PBHs with power law index $p$ ($p \neq 0$), maximum mass limit $M_{\rm max}$ and minimum mass limit $M_{\rm min}$ can be written as \cite{Carr:2017jsz, Chan:2020zry},
\begin{equation}
g(M_{\rm PBH})=\frac{p \rho_\odot}{M_{\rm max}^p - M_{\rm min}^p}M_{\rm PBH}^{p-1}\,\,,
\label{power}
\end{equation}
while the lognormal distribution with mean $\mu$ and standard deviation $\sigma$ is defined as \cite{Carr:2017jsz, Chan:2020zry},
\begin{equation}
g(M_{\rm PBH})=\frac{\rho_\odot}{\sqrt{2 \pi} \sigma M_{\rm PBH}}{\rm exp}\left(-\frac{{\rm ln}^2(M_{\rm PBH}/\mu)}{2 \sigma^2}\right)\,\,.
\label{log}
\end{equation}
The lognormal distribution of PBH mass function is first mentioned in Ref. \cite{Dolgov:1992pu} to discuss a mechanism of PBH formation for a model of baryogenesis which gives rise to large fluctuations (of baryon number) on small scales and small fluctuations on large scales. This lognormal distribution is a good approximation when it is considered that PBHs originate from a smooth symmetric peak in the inflationary power spectrum. For example in Ref. \cite{Kannike:2017bxn} it is shown that for single field double inflation theory, the associated PBH mass function is approximately lognormal. Numerical demonstration of this type of mass functions of PBH are given in Ref. \cite{Green:2016xgy}.

On the other hand, power law mass function of PBHs results from scale invariant density fluctuations or from the collapse of cosmic strings \cite{Carr:2017jsz}. In both the cases $p=-\frac{2 \omega}{1+\omega}$ where $\omega$ describes the equation of state at the time of PBH formation \cite{Carr:1975qj}. Since  PBHs form after inflation due to the inflation - generated density fluctuations and for the non inflationary Universe $\omega \in \left\lbrace -1/3,1\right\rbrace $, the range of power index $p$ would be $p \in \left\lbrace -1,1\right\rbrace $. Therefore, in this work two values $p$, namely $p=-0.5$ and $p=0.5$ are considered.
\subsection{Magnetic Field and Dark Matter Halo Profile}
In this case, the assumption has been made that the magnetic field near the Galactic Centre can be determined using the form \cite{Laha:2012fg, Chan:2020zry},
\begin{eqnarray}
  && B_0 \left( \frac{r_c}{r}\right)^{5/4} \hspace{5mm} {\rm for} \hspace{5mm} r\leqslant r_c \nonumber \\ 
B = && \label{magfield}\\
&& B_0 \left( \frac{r_c}{r}\right)^{2} \hspace{5mm} {\rm for} \hspace{5mm} r > r_c \,\,.\nonumber 
\end{eqnarray}
Here, $r_c$ is the accretion radius of the supermassive black hole (SMBH) at the central region and $r_c=0.04$ pc. While the strength of the magnetic field $B_0$ is assumed to be equal to 7.2 mG.

For DM halo profile, we have primarily adopted the Navarro - Frenk - White or NFW profile of DM halo to compute the results. The NFW profile can be written  as \cite{Navarro:1996gj},
\begin{equation}
\rho(r)=\frac{\rho_s}{\frac{r}{r_s}\left(1+\frac{r}{r_s}\right)^2}\,\,,
\label{NFW}
\end{equation}
where $r_s$ and $\rho_s$ are scale radius and scale density respectively. We have taken $r_s=20$ kpc and have chosen $\rho_s$ in such a way that it can provide the local DM density $\rho_\odot = 0.4$ GeV cm$^{ −3}$ \cite{Salucci:2010qr, Catena:2009mf} at a distance
$r = 8.5$ kpc from the Galactic Centre.

\section{Calculations and Results}\label{result}
In this section, using the formalism discussed in Sect. \ref{formalism}, we compute the synchrotron flux and IC flux for five different cases related to PBH decay and decay/annihilation of dark matter. These five cases are i) decay of PBH with monochromatic mass distribution, ii) decay of PBH with power law mass distribution, iii) decay of PBH with lognormal mass distribution iv) decay of DM and v) annihilation of DM. We then compare the computed results for all these five cases. Further, the detectability of such fluxes in the present and upcoming radio telescopes are also probed by comparing these with and bounds on the model parameters are obtained from the given SKA sensitivity \cite{SKA2017} and Jodrell Bank data \cite{10.1093/mnras/177.2.319}. Variations of the fluxes with the distance from the Galactic Centre are also calculated. Each of the cases is addressed for two ranges of $e^-/ e^+$ energy; high enrgy (order of GeV) $e^-/ e^+$ and low energy (order of MeV) $e^-/ e^+$. 
\subsection{Synchrotron and IC Radiation From the Produced High Energy Electrons/Positrons}\label{sect3.1}
\begin{figure}
\centering
\includegraphics[width=8cm,height=5cm]{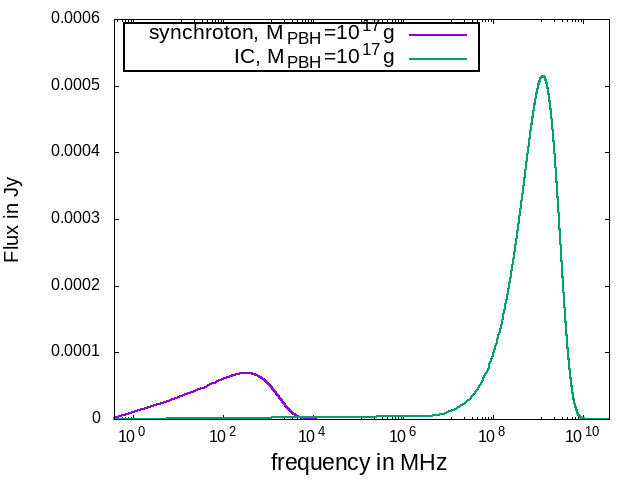}
\includegraphics[width=8cm,height=5cm]{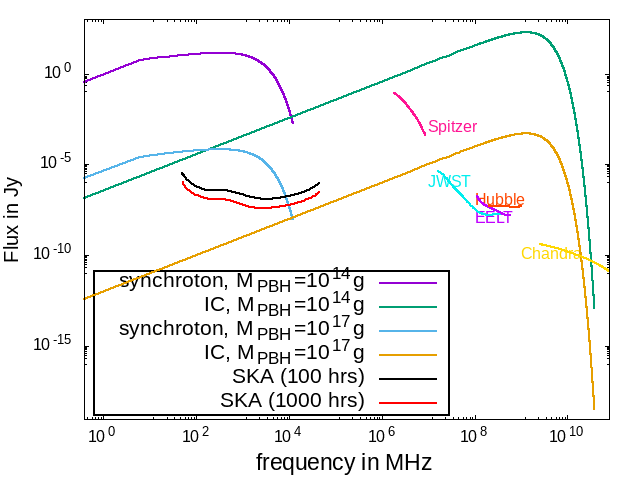}

\caption{{\it Left Panel:} Variation of the flux with frequency $\nu$ for synchrotron radiation (the purple line in the plot) and for IC radiation (the green line in the plot), originated due to the $e^-/ e^+$ evaporation of PBH with mass $M_{\rm PBH}$=$10^{17}$g.
 {\it Right Panel:} Variation of the flux for synchrotron radiation and for IC radiation with frequency $\nu$, generated due to the $e^-/ e^+$ evaporation of PBH with masses $M_{\rm PBH}$=$10^{17}$g and 4$\times$ $10^{14}$g. Sensitivity of SKA for 100 hrs and 1000 hrs are shown by the black and red curves respectively. Sensitivity of experiments in higher frequencies like Spitzer, JWST, EELT, Hubble, Chandra are also shown.
} 
\label{fig:mono}
\end{figure}

\begin{figure}
\centering
\includegraphics[width=8cm,height=5cm]{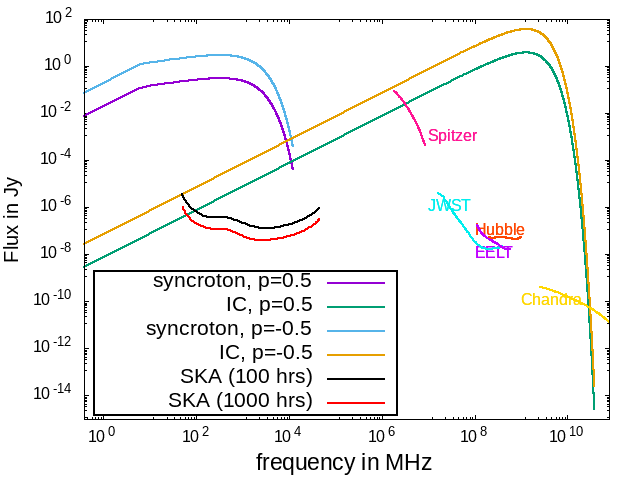}
\includegraphics[width=8cm,height=5cm]{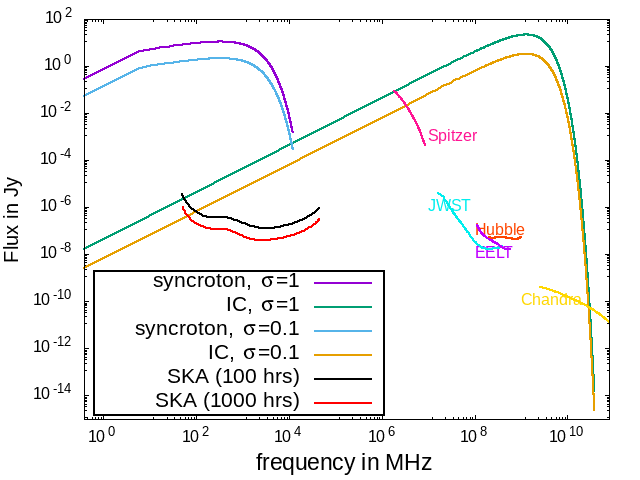}

\caption{{\it Left Panel:} Variation of the flux for synchrotron radiation and for IC radiation with frequency $\nu$, originated due to the evaporation of PBH with power law distribution of PBH mass where $M_{\rm min}$=4$\times 10^{14}$g, $M_{\rm max}$=$10^{17}$g and $p=-0.5, 0.5$.
 {\it Right Panel:} Variation of the flux for synchrotron radiation and for IC radiation with frequency $\nu$, originated due to the evaporation of PBH with lognormal distribution of PBH mass where average of the mass distribution $\mu=5 \times 10^{15}$g and variance $\sigma=0.1$ and $\sigma=1$. Sensitivity of SKA for 100 hrs and 1000 hrs are shown by the black and red curves respectively. Sensitivity of experiments in higher frequencies such as Spitzer, JWST, EELT, Hubble, Chandra are also shown.
} 
\label{fig:powerlog}
\end{figure}

\begin{figure}
\centering
\includegraphics[width=8cm,height=5cm]{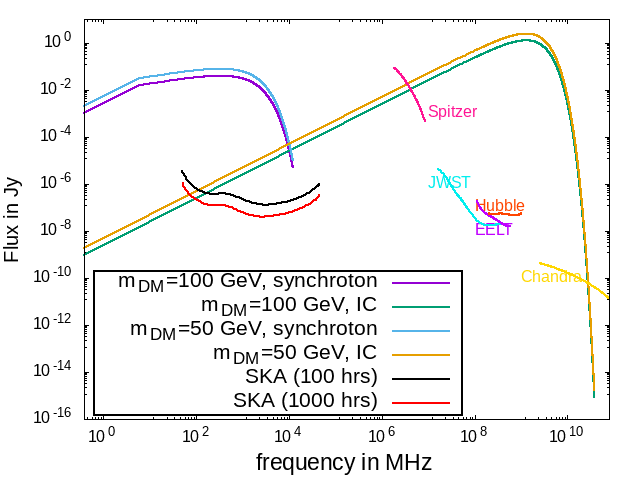}
\includegraphics[width=8cm,height=5cm]{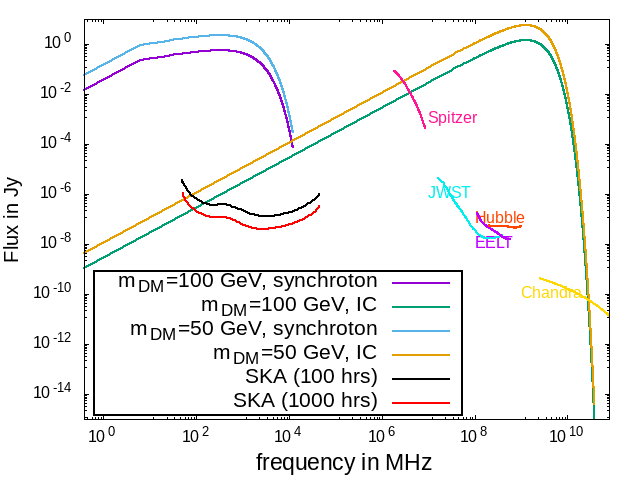}

\caption{{\it Left Panel:} Variation of the flux for synchrotron radiation and for IC radiation with frequency $\nu$, originated due to the decay of DM with mass $m_{\rm DM}$=50 GeV, 100 GeV and decay rate $\Gamma = 10^{-25}$s$^{-1}$. 
{\it Right Panel:} Variation of the flux for synchrotron radiation and for IC radiation with frequency $\nu$, originated due to the annihilation of DM with mass $m_{\rm DM}$=50 GeV and 100 GeV and annihilation rate $\left\langle   \sigma v \right\rangle = 10^{-26} {\rm cm}^3$s$^{-1}$. Sensitivity of SKA for 100 hrs and 1000 hrs are shown by the black and red curves respectively. Sensitivity of experiments in higher frequency domains such as Spitzer, JWST, EELT, Hubble, Chandra are also shown.
} 
\label{fig:DM}
\end{figure}

\begin{figure}
\centering
\includegraphics[width=8cm,height=5cm]{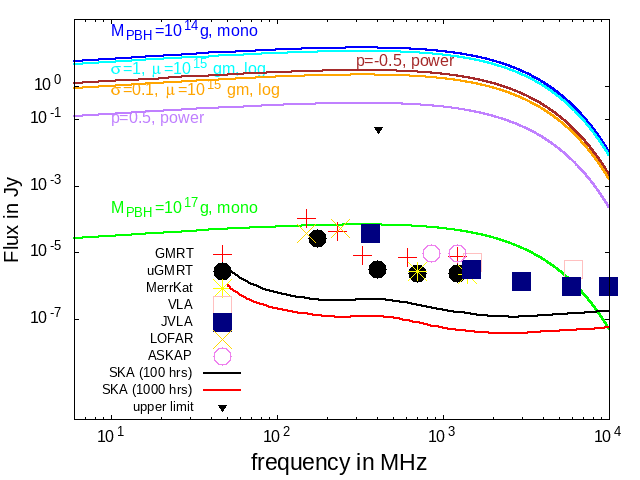}
\includegraphics[width=8cm,height=5cm]{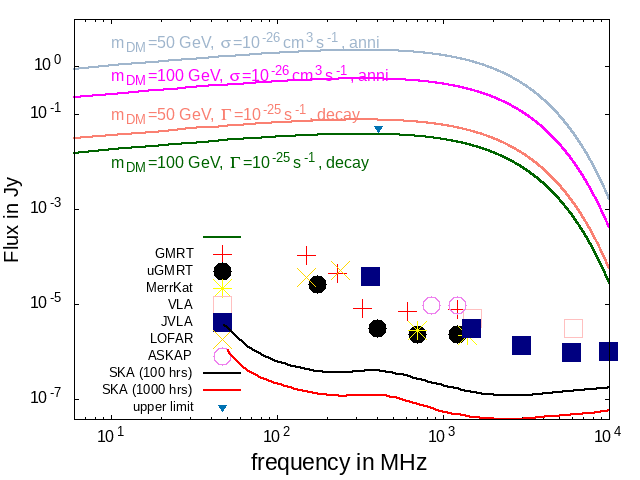}
\caption{{\it Left panel:} Variation of synchrotron flux with frequency generated from the evaporation of PBH with different mass values and different mass distributions.{\it Right panel:} Variation of synchrotron flux with frequency generated from the decay of DM and annihilation of DM with different masses.
The fluxes are compared with the sensitivities of different radio frequency experiments i.e., GMRT, uGMRT, MeerKat, VLA, JVLA, LOFAR, ASKAP, SKA (100 hrs), SKA (1000 hrs). These data are also shown in both the left and right panels.}
\label{fig:exp}
\end{figure}

In this scenario of high energy $e^-/e^+$, the synchrotron peak frequency and IC peak frequency are computed for electrons/positrons with energy $E\sim 0.1$ GeV. For these calculations, unless mentioned otherwise, $f_{\rm PBH}=10^{-7}$ and NFW profile of DM Halo have been adopted. The results are plotted in Figs. (\ref{fig:mono} - \ref{fig:exp}). 
It can be seen from Figs. \ref{fig:mono}- \ref{fig:exp} that in this case synchrotron emissions give rise to radio signals with the peak frequency at $\sim 338$ MHz while for the IC radiations ultraviolet to X-ray signals are obtained with the peak frequency at $\sim 1.2 \times 10^9$ MHz. The values of the synchrotron peak frequency depend on the energy of the electrons (positrons) and on the strength of the magnetic field while the IC peak frequencies depend on the energy of electrons (positrons) and photons. The sensitivities of SKA, Spitzer, JWST, EELT, Hubble and Chandra telescopes are also plotted for comparison.

From Fig. \ref{fig:mono} one notes that the intensities of the signals (for both the cases, i.e., synchrotron and IC) depend on the mass of the PBHs and the signal strengths  are higher for $M_{\rm PBH}=4\times 10^{14}$g than for $M_{\rm PBH}=10^{17}$g. This indicates that PBHs with smaller mass (or with higher temperature) evaporate in a higher rate than a PBH with larger mass. Moreover, it can be observed from Fig. \ref{fig:mono} that for both the values of PBH mass ($4\times 10^{14}$g and $10^{17}$g) the synchrotron radiations are in the detectable range of SKA data but the detectability of the IC fluxes are observed on the basis of the experiments in higher frequency like Spitzer, JWST, EELT, Hubble and Chandra.     

In Fig. \ref{fig:powerlog}, variations of the flux for synchrotron radiations and IC radiations with frequency $\nu$, originated due to the evaporation of PBH with power law mass distribution (left panel) and lognormal mass distribution (right panel) are shown. For the power law distribution two values of $p$, namely $p=-0.5, $ $0.5$, are considered with PBH masses within the range $M_{\rm min}$=4$\times 10^{14}$g to $M_{\rm max}$=$10^{17}$g. From Fig. \ref{fig:powerlog}, it is observed that both the synchrotron and IC radiation fluxes are higher when the power index $p$ of the PBH power law mass distribution is taken to be $p=-0.5$ than when $p=0.5$ is considered.
It can be noted from Eq. \ref{power} that for $p=-0.5$ the $g(M_{\rm PBH}) \propto \frac{1}{M_{\rm PBH}^{3/2}}$ and for $p=0.5$ the $g(M_{\rm PBH}) \propto \frac{1}{\sqrt{M_{\rm PBH}}}$ respectively. 
Hence, for the case of $p=-0.5$ the weight factor of PBH mass function $g(M_{\rm PBH})$ increases more rapidly with smaller values of $M_{\rm PBH}$ than the case when $p=0.5$ and as PBHs with smaller mass values provide larger $e^-/ e^+$ fluxes the flux densities with $p=-0.5$ are greater than the same with $p=0.5$.


The right panel of Fig. \ref{fig:powerlog} shows similar results but for lognormal distribution of PBH masses with $\mu=5 \times 10^{15}$ g and for two values of the variance, namely $\sigma=0.1$ and $1$. It can be noted from Fig. \ref{fig:powerlog} that a larger variance ($\sigma=1$) yields larger fluxes for synchrotron radiations and IC radiations. This is however expected since a larger value of variance $\sigma$ includes the contribution of larger range of PBH masses to the flux densities. This may be understood that larger the mass ranges of PBHs smaller are the range of values of $M_{\rm PBH}$ (with larger evaporation rate) that can contribute to the fluxes. Hence the flux densities for $\sigma=1$ are larger than the flux densities when $\sigma =0.1$ is adopted. Similar comments as in Fig. \ref{fig:mono} in relation to the peak frequencies and detectability of the signals for SKA data and Spitzer, JWST, EELT, Hubble, Chandra data can also be made from Fig: \ref{fig:powerlog}.

We now calculate the fluxes for synchrotron and IC radiations that could be originated from the possible decay products ($e^+, e^-$) of DM. For demonstrative purpose we consider two values of masses of the decaying DM. The results are plotted in left panel of Fig. \ref{fig:DM}. In Fig. \ref{fig:DM} (left panel) variations of the flux with frequency $\nu$ are plotted for synchrotron radiations and IC radiations originated due to the decay of DM with mass $m_{\rm DM} = 50$ GeV and 100 GeV and decay rate $\Gamma = 10^{-25}$ s$^{-1}$. In the present calculations flux density is always higher when decaying DM mass is 50 GeV than when the same is 100 GeV. This is expected since from Eq. \ref{Qdecay} it can be understood that smaller values of DM mass will lead to higher values of flux densities. The flux density also depends on the decay rate (larger the decay rate, larger would be the flux density). To this end, later in this work (in section \ref{bound}) we obtain bounds on the DM decay rate and DM mass in the present analysis using the experimental results.

Similar computations are made for the synchrotron and IC fluxes with the consideration that these fluxes are originated due to the motion of the products of the DM annihilation in Galactic magnetic field and the scattering of these products. The fluxes are obtained for two annihilating DM masses, namely 50 GeV and 100 GeV with annihilation cross section $\left\langle   \sigma v \right\rangle = 10^{-26} {\rm cm}^3$s$^{-1}$. The results are plotted in the right panel of Fig. \ref{fig:DM}. 
It can again be noted from the figure that values of flux density depend on DM mass and it is higher for smaller value of DM mass (i.e., 50 GeV). This can be explained from Eq. \ref{Qdecay} that smaller values of DM mass would lead to higher values of the flux density. Moreover, the values of the flux density also depend on the annihilation rate of DM (larger annihilation rates provide larger flux densities). In section \ref{bound} we discuss the bounds on the annihilation rates and DM mass on the basis of the experimental results.
For both the plots (left panel and right panel) of Fig. \ref{fig:DM} the flux densities are compared with experimental results of SKA and Spitzer, JWST, EELT, Hubble, Chandra.  

In this high energy ($\sim 0.1$ GeV) $e^-/ e^+$ case, we have computed that the synchrotron emissions induced by the decay products ($e^-/ e^+$) of PBH or the decay/annihilation products ($e^-/ e^+$) of DM produce radio waves. Therefore in Fig. \ref{fig:exp}, the synchrotron flux densities are compared with the present and future radio telescopes' data.

A versatile next generation radio telescope to observe a large area of the sky is Square Kilometre Array (SKA) \cite{SKA2017, Kar:2019cqo, Bertolami:2018lel, Kar:2018rlm}. It operates in the frequency range of 70 MHz to 10 GHz and predicts the flux density limit even in the order of $\mu$Jy. 

Another low frequency radio telescope is Giant Metrewave Radio Telescope (GMRT) \cite{uGMRT2017} located in India. It operates in the frequency range of 150 - 1500 MHz with
five discrete bands namely 130 - 170 MHz, 225 - 245 MHz, 300 - 360 MHz, 580 - 600 MHz, 1000 - 1450 MHz and the r.m.s sensitivities of these five discrete bands are
0.7, 0.25, 0.04, 0.02, 0.03 respectively in the mJy unit. The upgraded version of GMRT, the uGMRT \cite{uGMRT2017} is expected to be sensitive in the frequency ranges 130 - 260 MHz, 250 - 500 MHz, 550 - 900 MHz,
1000 - 1450 MHz and it is more sensitive in the low frequency band ranges of 250 to 1000 MHz. 
In addition a new radio telescope MerrKAT \cite{Brederode:2018gh}, a precursor to SKA telescope, will be operated in the frequency bands 0.7 - 2.5 GHz, 0.7- 10 GHz. A low frequency radio interferometer is the LOw-Frequency ARray or LOFAR \cite{van_Haarlem_2013} is designed to cover the low-frequency range from 10 - 240 MHz. The Jansky Very Large Array (JVLA) \cite{JVLA} is another new generation radio telescope that operates between 1 - 50 GHz while the Very Large Array or the VLA operates around 1.4 GHz and 5 GHz \cite{VLA}. Operational frequency range of another upcoming radio telescope ASKAP is 0.7 GHz - 1.8 GHz \cite{SKA2017}. The radio telescope Jodrell Bank \cite{10.1093/mnras/177.2.319} measures radio flux within 4 arcsec at the Galactic Centre and it measurement reveals that at 408 MHz the upper bound on the radio flux is 50 mJy.


In Fig. \ref{fig:exp}, we have compared all our results for the synchrotron flux densities with the data/sensitivity of the above mentioned radio telescopes. In the left panel of Fig. \ref{fig:exp} flux densities due to the decay of PBHs with different masses and mass distributions are compared with the experimental data while in the right panel of Fig. \ref{fig:exp} similar comparisons have been made for the case of DM decay/annihilation products. It can be noted from the figure that DM annihilation products yield a larger flux than DM decay. A point corresponding to ``upper limit'' shown in Fig. \ref{fig:exp} indicates that upper limit of the radio flux densities at the Galactic Centre at frequency 408 MHz is 50 mJy. This provides tight constraints on the parameter $f_{\rm PBH}$ of PBH distributions and DM annihilation cross sections or decay rates as well as on the masses of PBH and DM. The bounds on these quantities from experimental results will be discussed later in section \ref{bound}.

\subsubsection{Variation of the Flux with Distance from the Galactic Centre}
This can be noted from the formalism in Sect. \ref{formalism} that the synchrotron and IC flux densities would vary with the distance $r$ from the Galactic Centre. The variations are principally caused due to the variations of magnetic field $B(r)$ with $r$ and due to the variations of DM densities with $r$. In this section we compute synchrotron flux densities at different distances ($r$) from the Galactic Centre for the cases considered earlier i.e., PBH evaporation and DM annihilation/decay. In this work we have considered a region within 2 pc (4 arcsec) from the Galactic Centre.

\begin{figure}
\centering
\includegraphics[width=8cm,height=5cm]{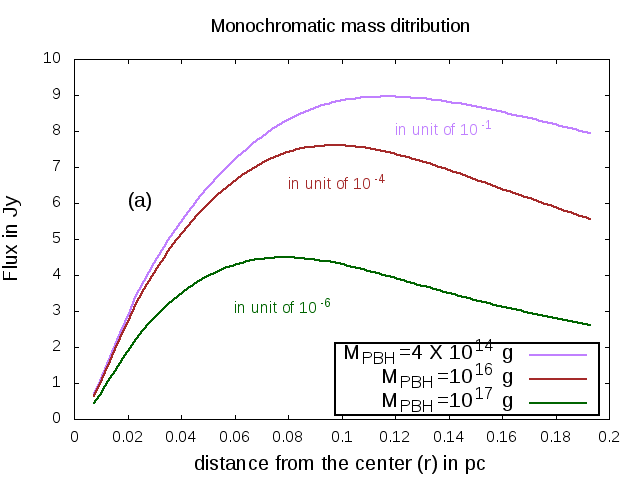}
\includegraphics[width=8cm,height=5cm]{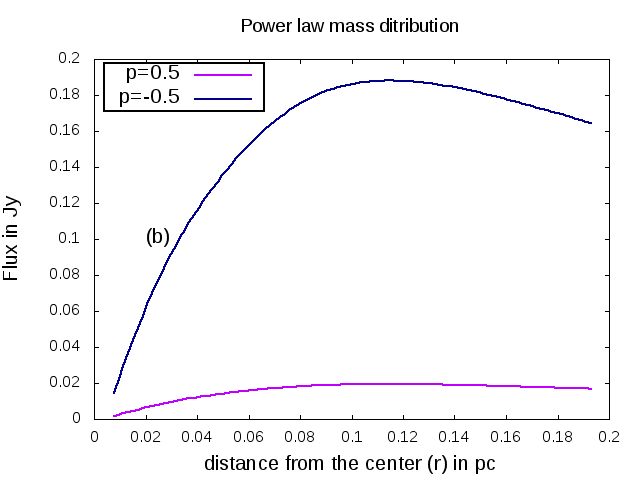}
\includegraphics[width=8cm,height=5cm]{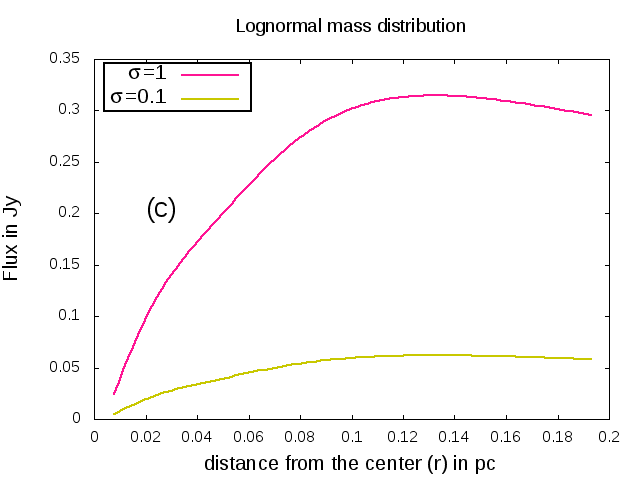}
\caption{(a) Variations of the synchrotron flux densities with the distance from the Galactic Centre are plotted due to the evaporation of PBHs with monochromatic mass distributions. Variations are plotted for three PBH masses $4\times 10^{14}$ g, $10^{16}$ g and $10^{17}$ g. (b) Variations of the synchrotron flux densities with the distance from the Galactic Centre are plotted due to the evaporation of PBH with power law mass distributions for two values of $p$, namely -0.5 and 0.5. (c) Variations of the synchrotron flux densities with the distance from the Galactic Centre are plotted due to the evaporation of PBH with lognormal mass distributions for two values of $\sigma$, namely 0.1 and 1.}
\label{vary:PBH}
\end{figure}

Fig. \ref{vary:PBH}(a) shows variations of the synchrotron flux densities with the distance from the Galactic Centre when the evaporation of PBHs with monochromatic mass distribution is considered. Variations are plotted for three PBH masses namely, $4\times 10^{14}$g, $10^{16}$g and $10^{17}$g. It can be noted that for each of these cases the flux density shows a bell shaped curve i.e., it increases with the distance from the Galactic Centre until it achieves the largest flux density at a fixed distance and then it starts to decrease. But the position of a peak depends on the PBH mass values. For larger values of $M_{\rm PBH}$ the flux densities attain their peak value at smaller distances $r$. It can be seen from Fig. \ref{vary:PBH}(a) that the positions of the highest flux densities for $M_{\rm PBH}=4\times 10^{14} $g, $10^{16}$g and $10^{17}$g are at $\sim 0.11$, $\sim 0.09$ and $\sim 0.08$ respectively. Therefore, the positions of such peaks could be a possible probe to determine the value of the PBH mass.
In order to compare the peak positions of the flux densities for three different values of $M_{\rm PBH}$, we have plotted the fluxes for $M_{\rm PBH}=4\times 10^{14} $g, $10^{16}$g and $10^{17}$g in the units of $10^{-1}$, $10^{-4}$ and $10^{-6}$ Jy respectively. From Fig. \ref{vary:PBH}(a) one finds that the peak values of the flux density for these three cases are $\sim$ $9 \times 10^{-1}$ Jy, $7 \times 10^{-4}$ Jy and $4 \times 10^{-6}$ Jy respectively.

Moreover, it is also observed from Fig. \ref{vary:PBH} that larger the PBH masses (corresponds to smaller flux densities) nearer are the distances $r$ from the Galactic Centre at which the synchrotron radiation peaks. Therefore, the different peak positions for different PBH masses indicate that for larger masses of PBH, the flux decreases with $r$ much faster than the case for smaller PBH masses. While for the case of heavier PBHs, significant synchrotron flux may not be obtained at larger distances from the Galactic Centre but for smaller PBH masses the chances to obtain detectable PBH induced flux at extended region are larger. It indicates that more massive the PBH is more probable they are to be present closer to the Centre of the Galaxy and therefore, the synchrotron flux caused by the evaporation of PBH in and around the Galactic Centre region is likely to attempt its peak value at a distance closer to the Centre of the Galaxy. Moreover, the shape of the curves in Fig. \ref{vary:PBH} can be understood from the form of the magnetic field $B(r)$ given in Eq. \ref{magfield}, the shape of DM halo profile $\rho(r)$ (Eq. \ref{NFW}) and total energy loss term $b(E,r)$ (Eq. \ref{loss}). The $r$ dependence of synchrotron flux density can be approximately of the form of $\sim \frac{1}{B(r)} \rho(r)$ giving rise to such bell shaped curve on integration over energy.



In Fig. \ref{vary:PBH}(b), similar plots as in Fig. \ref{vary:PBH}(a) are shown but for the power law mass distribution of the PBHs. Similar bell shaped curves are obtained with different peak positions depending upon the values of the power index $p$ (here we have considered $M_{\rm min} = 4 \times 10^{14}$g and $M_{\rm max} = 10^{17}$g). For $p=-0.5$, maximum flux density ($\sim 0.18$ Jy) is obtained at $r\sim 0.12$ pc while for $p=0.5$, the maximum value ($\sim 19 \times 10^{-3}$ Jy) is obtained at $r\sim 0.1$ pc. Here also it can be observed that for $p=-0.5$ (which corresponds to larger flux density), the peak position of the flux density is located at a distance $r$ from the Galactic Centre which is further from the peak value of lower synchrotron flux when $p=0.5$ is adopted for power law distribution of PBH masses. It is already discussed that for $p=-0.5$ case, there are more contributions from smaller masses of PBH and hence significant flux densities at larger distances from the Galactic Centre can be obtained.

In Fig. \ref{vary:PBH}(c) variations of synchrotron flux densities at different positions from the Galactic Centre for the case of lognormal mass distribution of PBHs are computed and plotted. The lognormal distribution of PBHs is considered for this computation for mean $\mu=5 \times 10^{15}$g and two values of variance $\sigma$, namely $\sigma=0.1$ and $\sigma = 1$. It can be seen from the Fig. \ref{fig:powerlog} that $\sigma=1$ yields larger flux densities. It is noted from Fig. \ref{vary:PBH}(c) that for $\sigma=1$ the peak position of the flux density is obtained at a larger distance $r$ than the flux density calculated with $\sigma=0.1$.

\begin{figure}
\centering
\includegraphics[width=8cm,height=5cm]{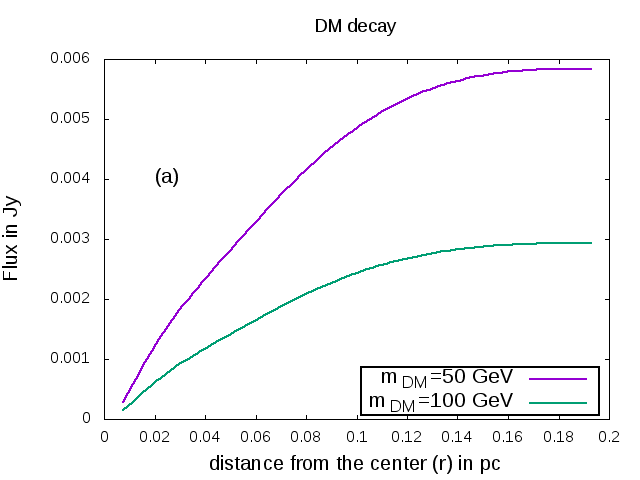}
\includegraphics[width=8cm,height=5cm]{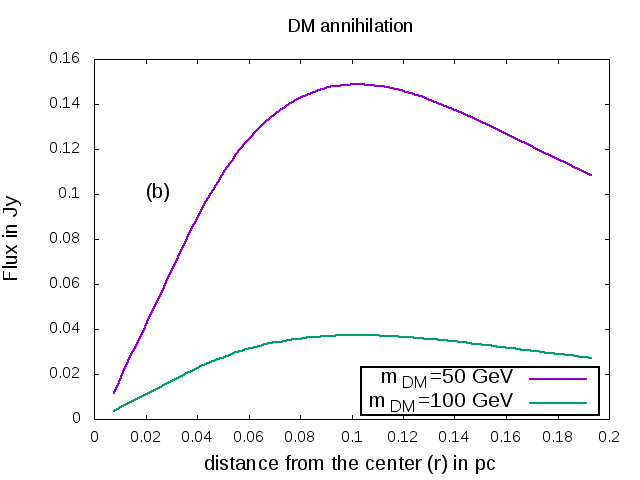}
\caption{(a) The variations of the synchrotron flux densities from the Galactic Centre are plotted for the decay of DM. (b) The variations of the flux densities from the Galactic Centre are plotted for the annihilation of DM.
For both the cases flux densities are compared for two mass values of DM, namely $m_{\rm DM}=50$ GeV, 100 GeV.}
\label{vary:DM}
\end{figure}

In Fig. \ref{vary:DM} the variations of the synchrotron flux densities with distance from the Galactic Centre are plotted for the decay/annihilation of DM. The results for the decay of DM and the annihilation of DM are plotted in Fig. \ref{vary:DM}(a) and Fig. \ref{vary:DM}(b) respectively. For both the cases, flux densities are computed for two mass values ($m_{\rm DM}$) of DM namely, 50 GeV and 100 GeV. It can be noted from the figure that the peak positions of the flux densities depend on the DM mass. The position of the peak flux densities from the Galactic Centre could be a useful probe to determine the mass of the DM at the Galactic Centre. The plots in Fig. \ref{vary:DM} are obtained by choosing fixed values of decay rate $\Gamma=10^{-25}$ s$^{-1}$ and annihilation cross section $\left\langle   \sigma v \right\rangle=10^{-26}$ cm$^3$ s$^{-1}$. It can also be observed from the figure that as 100 GeV DM produces lower flux densities than 50 GeV DM (both for the decay or the annihilation), the peaks of flux densities when decay or annihilation of 100 GeV DM is considered, occur nearer to the Galactic Centre i.e., at a smaller distance $r$ than the case when DM mass is considered to be 50 GeV.

\begin{figure}
\centering
\includegraphics[width=8cm,height=5cm]{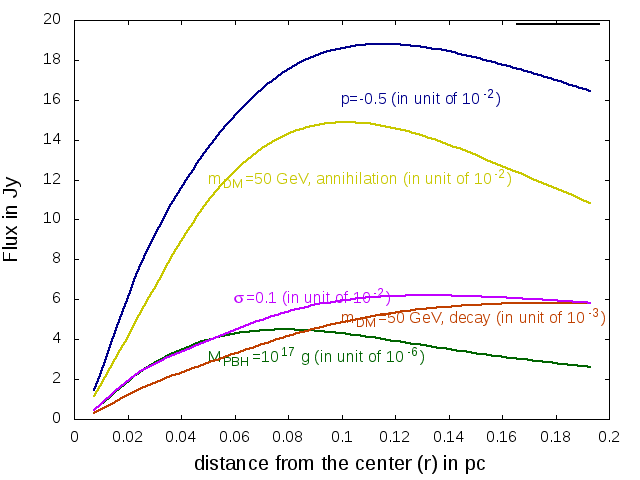}
\caption{Variations of the synchrotron flux with $r$ are compared for four cases, decay of PBH with $M_{\rm PBH}=10^{17}$ g, decay of PBH with power law distribution with $p=-0.5$, decay of PBH with lognormal mass distribution with $\sigma =0.1$, decay of 50 GeV DM, annihilation of 50 GeV DM.}
\label{vary:comparison}
\end{figure}

In Fig. \ref{vary:comparison} variations of the synchrotron flux with $r$ are compared for five cases considered so far namely, decay of PBH with $M_{\rm PBH}=10^{17}$g (the dark green line in the figure), decay of PBH with power law distribution for $p=-0.5$ (the dark blue line in the figure), decay of PBH with lognormal mass distribution for the case $\sigma = 0.1$ (the magenta line in the figure), decay of 50 GeV DM (the brown line in the figure) and annihilation of DM (the yellow line in Fig. \ref{vary:comparison}). It is to be understood that all the plots in Fig. \ref{vary:comparison} have already been showm in Fig. \ref{vary:PBH} - \ref{vary:DM} but here they are plotted together for a direct comparison. In order to compare these five cases, the flux densities are plotted in the units of $10^{-6}$, $10^{-2}$, $10^{-2}$, $10^{-3}$ and $10^{-2}$ Jy respectively. It can again be observed (from Fig. \ref{vary:comparison}) that different cases show different peak values and peak positions of synchrotron flux densities. Thus such peak values and peak positions could be a possible probe to determine the mass of PBH or the index parameters of the PBH mass distributions and mass of DM as well.

\begin{figure}
\centering
\includegraphics[width=8cm,height=5cm]{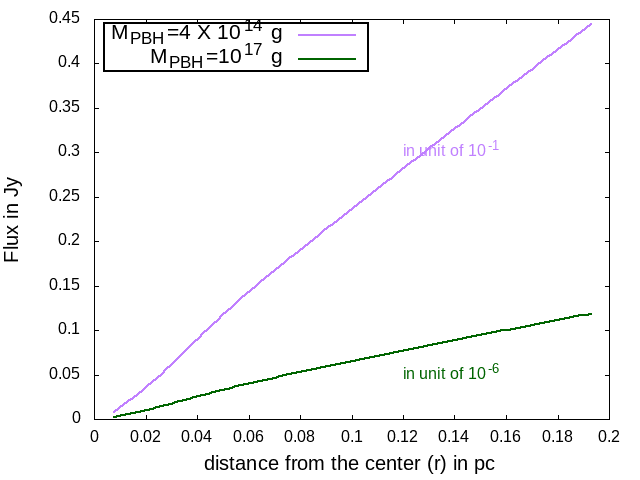}
\caption{Variations of the IC flux densities, originated due to the evaporation of PBH with monochromatic mass distribution, with the distance from the Galactic Centre.}
\label{vary:IC}
\end{figure}

Now, we focus on IC flux densities for the case of PBH evaporations and decay and annihilations of DM similar to what considered for the synchrotron case. First we consider the variations of IC flux densities, originated due to the evaporation of PBH with monochromatic mass distribution, with the distance $r$ from the Galactic Centre. This is shown in Fig. \ref{vary:IC}. From Fig. \ref{vary:IC} it can be seen that IC flux density increases with $r$ since the IC flux density approximately varies with distance $r$ as $\sim \frac{1}{B(r)^2} \rho(r)$ (note that this is different from the dependence of synchrotron flux density on $r$ ($\sim \frac{1}{B(r)} \rho(r)$)). Hence, the variations of synchrotron flux densities and IC flux densities with $r$ show different behaviour. Variations of IC flux densities with $r$ for the cases of power law or lognormal mass distributions of PBH or annihilation/decay of DM are also computed and similar behaviours of flux densities (as in Fig. \ref{vary:IC}) are obtained. These are however not shown here.

\begin{figure}
\centering
\includegraphics[width=8cm,height=5cm]{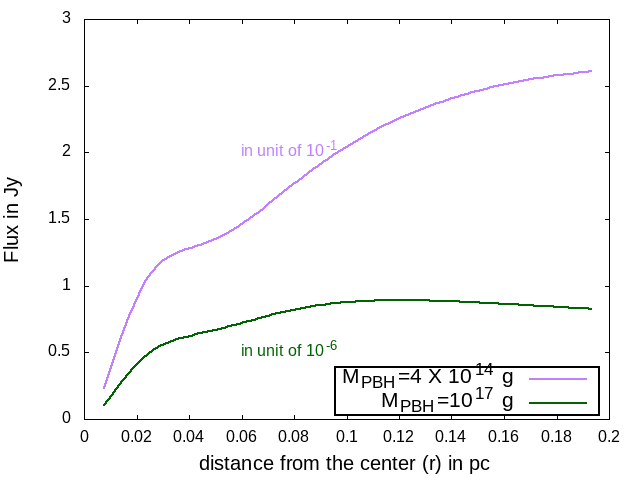}
\caption{Variations of the synchrotron flux densities, originated due to the evaporation of PBH with monochromatic mass distribution ($M_{\rm PBH}=4 \times 10^{14}$ g, $10^{17}$ g), with the distance from the Galactic Centre with EINASTO DM halo profile.}
\label{vary:einasto}
\end{figure}

The computations for Figs. \ref{vary:PBH} - \ref{vary:IC} are performed by adopting NFW profile for DM halo density $\rho(r)$ (Eq. \ref{NFW}). We now perform the same calculations but with a different DM density profile namely, EINASTO profile for $\rho(r)$. In order to demonstrate the dependence of flux density on DM halo profile, we have plotted in Fig. \ref{vary:einasto}, variations of flux density with $r$ for the case of PBH evaporation  with monochromatic mass distribution by considering EINASTO profile for DM halo density.  The EINASTO profile is given by \cite{Graham:2005xx},
\begin{equation}
\rho(r)=\frac{\rho_s}{{\rm exp}\left[\frac{2}{\alpha}\left(\left(\frac{r}{r_s}\right)^\alpha-1 \right)\right]}\,\,.
\end{equation}
For this calculations $\alpha=0.17$ has been adopted while other symbols have similar meaning as in Eq. \ref{NFW}.
The evaporation of PBH with monochromatic mass distribution with two PBH masses namely, $M_{\rm PBH}=4 \times 10^{14}$ g and $10^{17}$ g are considered and the results are plotted in Fig. \ref{vary:einasto}. By comparing Fig. \ref{vary:PBH} (a) and Fig. \ref{vary:einasto} it can be noted that the variations of the flux densities with $r$ are different for the two DM halo profiles (NFW and EINASTO). Hence, the peak positions and peak values of the flux densities depend on the profile of the DM halo.

\subsubsection{Constraints on the Model Parameters from Experimental Limits for High Energy Electrons and Positrons}\label{bound}
\begin{figure}
\centering
\includegraphics[width=8cm,height=5cm]{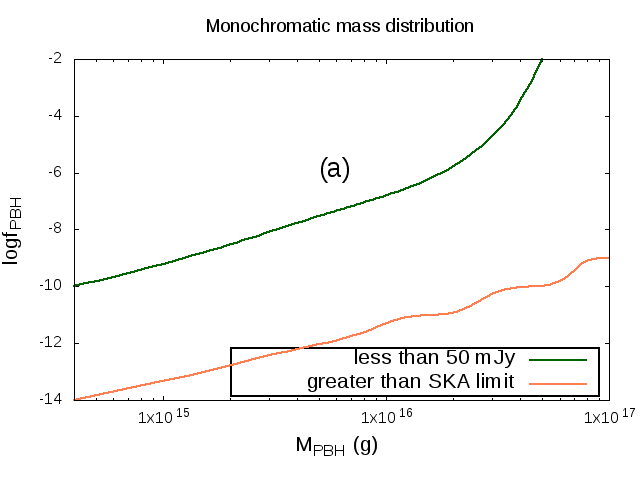}
\includegraphics[width=8cm,height=5cm]{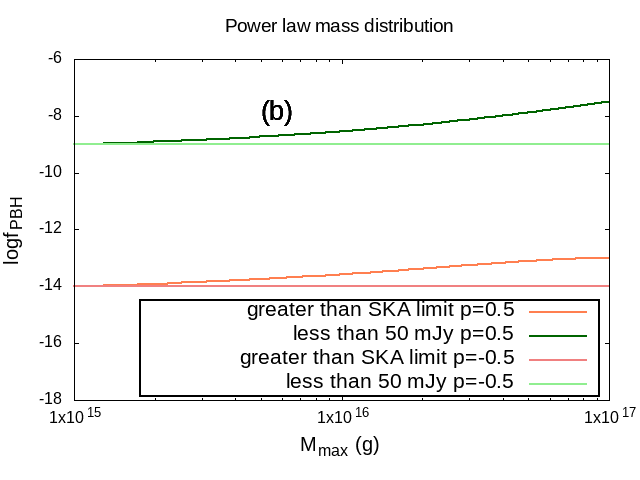}
\includegraphics[width=8cm,height=5cm]{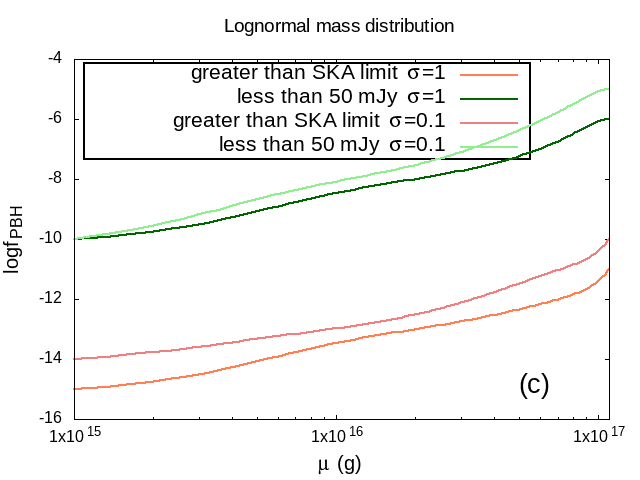}
\caption{(a) The upper limits and lower limit of $f_{\rm PBH}$ as a function of the PBH monochromatic mass. (b) The upper limits and lower limit of $f_{\rm PBH}$ as a function of the maximum mass in power law mass distribution of PBH. (c) The upper limits and lower limit of $f_{\rm PBH}$ as a function of the mean value of lognormal mass distribution of PBH.}
\label{limit_mono1}
\end{figure}

\begin{figure}
\centering
\includegraphics[width=7.5cm,height=4.5cm]{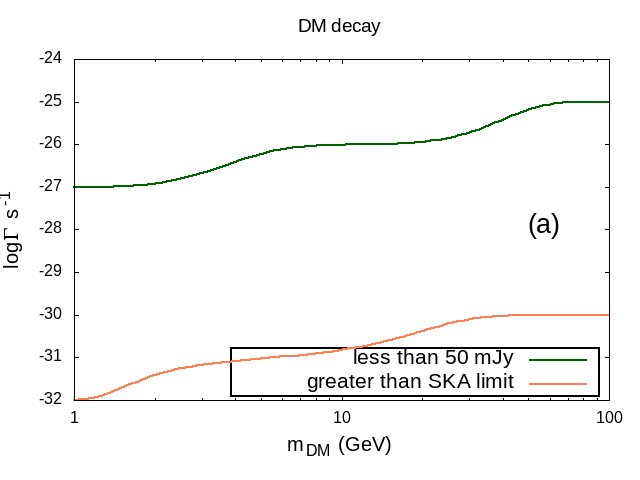}
\includegraphics[width=7.5cm,height=4.5cm]{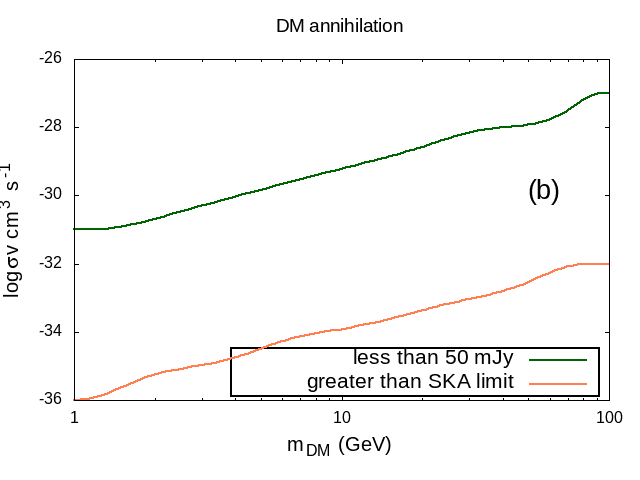}
\caption{(a) The upper limit and lower limit of $\Gamma$ as a function of the DM mass $m_{\rm DM}$ for the decay of DM. (b) The upper limit and lower limit of  $\left\langle   \sigma v \right\rangle$ as a function of the DM mass $m_{\rm DM}$ for the annihilation of DM.}
\label{limit_DM1}
\end{figure}

From the above discussions it may be evident that certain demonstrative values have been adopted for the quantities $f_{\rm PBH}$, DM decay constant $\Gamma$, DM annihilation cross section $\langle \sigma v \rangle$ in order to investigate the possible nature and variations of synchrotron radiations that could have been originated due to the influence of Galactic magnetic field on $e^+/e^-$ produced from the evaporation of PBH or from the decay/annihilation of DM. In this section we make an attempt to constrain these quantities or parameters by the SKA sensitivities and from the Jodrell Bank telescope's result that radio flux within 4 arcsec at the Galactic Centre should not be greater than 50 mJy at 408 MHz. We furnish our results in Fig. \ref{limit_mono1} and Fig. \ref{limit_DM1}. 

In Fig. \ref{limit_mono1}(a) the region above the orange line labeled as ``greater than SKA limit'' is for the values of $f_{\rm PBH}$ and $M_{\rm PBH}$ for which the intensity of the synchrotron radiation signal would be in the detectable range of SKA (i.e., the synchrotron flux density is above the order of $\mu$Jy). It can be observed from Fig. \ref{limit_mono1}(a) that for smaller values of $M_{\rm PBH}$, the allowed range for $f_{\rm PBH}$ is smaller than that for higher values of $M_{\rm PBH}$. This is expected because a PBH with smaller mass ($M_{\rm PBH}$) evaporates at a higher rate resulting in larger $e^-/ e^+$ fluxes and hence detectable radio signals through synchrotron emission even for smaller PBH densities (or $f_{\rm PBH}$) may be detected. From the figure it can be found that for $M_{\rm PBH}=10^{17}$ g, the minimum value of $f_{\rm PBH}$ is $10^{-9}$ while for $M_{\rm PBH}=4\times10^{14}$ g, it can probe the $f_{\rm PBH}$ down to $10^{-14}$. Moreover, in Fig. \ref{limit_mono1}(a) we have also shown the upper limits of $f_{\rm PBH}$ for different $M_{\rm PBH}$ that satisfy the highest limit of the flux (50 mJy) at 408 MHz. This is shown in Fig. \ref{limit_mono1}(a) by the dark green line in the plot labeled as ``less than 50 mJy''. Hence the region below the dark green line is allowed by Jodrell Bank's prediction. In Fig. \ref{limit_mono1}(b), we constrain the parameter $f_{\rm PBH}$, using SKA sensitivity, for the case of power law mass distribution of the PBH. To this end, $f_{\rm PBH}-M_{\rm max}$ parameter space ($M_{\rm max}$ is the maximum mass in the PBH distribution) is constrained for two values of power law index $p$ namely, $p=0.5$ and $p=-0.5$. The radio flux upper limit at 408 MHz for the power law mass distribution of PBH is also used. In Fig. \ref{limit_mono1}(b), the $M_{\rm min}$ (minimum value of PBH maa) is fixed at $4\times 10^{14}$g. It can be noted from Fig. \ref{limit_mono1}(b) that even for $p=0.5$ and $M_{\rm max}=10^{17}$g, $f_{\rm PBH}$ does not go beyond $\sim 10^{-8}$ in order to satisfy the experimental constraints but it can be as low as $\sim 10^{-14}$ to produce detectable radio signals in SKA. We then constrain $f_{\rm PBH}$ if the PBH masses follow a lognormal distribution. Here the parameter space is $f_{\rm PBH}-\mu$ is constrained for two values of $\sigma$ - the variance of the mass distribution. The results are plotted in Fig. \ref{limit_mono1}(c). 
In Fig. \ref{limit_mono1}(c) the region above the orange line labeled as ``greater than SKA limit'' describes the values of $f_{\rm PBH}$ to obtain detectable flux density in SKA and the dark green line labeled as ``less than 50 mJy'' indicates the upper limit of $f_{\rm PBH}$ to satisfy the highest limit of the flux (50 mJy) at 408 MHz for lognormal mass distribution of PBH. The bounds on $f_{\rm PBH}$ and the mean values of the mass distribution $\mu$ are plotted for $\sigma=0.1,$ 1. In this case $f_{\rm PBH}$ can be $\sim 10^{-5}$ for $\mu=10^{17}$ g, $\sigma=0.1$  and $f_{\rm PBH}\sim 10^{-6}$ for $\mu=10^{17}$ g, $\sigma=1$. Moreover, for $\mu=10^{15}$ g, SKA can detect the radio fluxes even for $f_{\rm PBH}$ as small as $\sim 10^{-14}$ ($\sigma=0.1$) and $\sim 10^{-15}$ ($\sigma=1$).

In this context it may be mentioned that the constraints on $f_{\rm PBH}$ as a function of mass of the PBH for monochromatic mass distribution or for the extended mass functions of the PBH are discussed in literature \cite{Carr:2017jsz}. Also, different other authors have used different cosmic-ray, gamma-ray or X-ray data \cite{Boudaud:2018hqb} to provide the upper limit of the $f_{\rm PBH}$ values. Moreover, in Ref. \cite{Carr:2020gox} the constraints are given for PBH mass range of $10^{13} - 10^{17}$ g with CMB
anisotropy damping limit and the Galactic positron limit. In this work we have provided another tight constraints on PBH mass function parameters with the help of SKA sensitivity and Jodrell Bank's data.
We now constrain the parameters, DM decay width $\Gamma$ and DM annihilation cross section $\langle \sigma v \rangle$, using the SKA limits. These are shown in Fig. \ref{limit_DM1}(a) and \ref{limit_DM1}(b). In Fig. \ref{limit_DM1}(a) the constrained parameter space $\Gamma - m_{\rm DM}$ ($m_{\rm DM}$ being the DM mass) is shown while in Fig. \ref{limit_DM1}(b) the DM annihilation cross section $\langle \sigma v \rangle$ is constrained for different values of $m_{\rm DM}$. In both Fig. \ref{limit_DM1}(a) and Fig. \ref{limit_DM1}(b) the regions above the orange line labeled as ``greater than SKA limit'' describe the values of the decay rate $\Gamma$ and annihilation cross section $\left\langle   \sigma v \right\rangle$ respectively for which the flux of the synchrotron radiation signal would be in the detectable range of SKA (flux density is above the order of $\mu$Jy). It can be seen from Fig. \ref{limit_DM1}(a) that the allowed values of $\Gamma$ are smaller for smaller $m_{\rm DM}$ (minimum value of $\Gamma$ is $10^{-32}$ s$^{-1}$ for $m_{\rm DM}= 1$ GeV while minimum value of $\Gamma$ is $10^{-30}$ s$^{-1}$ for $m_{\rm DM}= 100$ GeV). Similarly in Fig. \ref{limit_DM1}(b), $\left\langle   \sigma v \right\rangle$ is smaller for smaller $m_{\rm DM}$ in order to satisfy the SKA limit (minimum value of $\left\langle   \sigma v \right\rangle$ is $10^{-36}$ cm$^3$s$^{-1}$ for $m_{\rm DM}= 1$ GeV while minimum value of $\left\langle   \sigma v \right\rangle$ is $10^{-32}$ cm$^3$s$^{-1}$  for $m_{\rm DM}= 100$ GeV). This is also expected from Fig. \ref{fig:exp}. As larger is the DM mass smaller is the flux densities, larger decay rates or larger annihilation cross sections are required to satisfy the detectable criteria of SKA in this situations. Tight constraints on the upper limits of $\Gamma$ and $\left\langle   \sigma v \right\rangle$ arise from the fact that radio flux densities can not be grater than 50 mJy at 408 MHz within 2 pc of the Galactic Centre. Thus the dark green lines in Fig. \ref{limit_DM1} labeled ``less than 50 mJy'' represent the upper limits.
\subsection{Synchrotron Radiation and IC Radiation from the Produced Low Energy Electrons/Positrons}

\begin{figure}
\centering
\includegraphics[width=8cm,height=5cm]{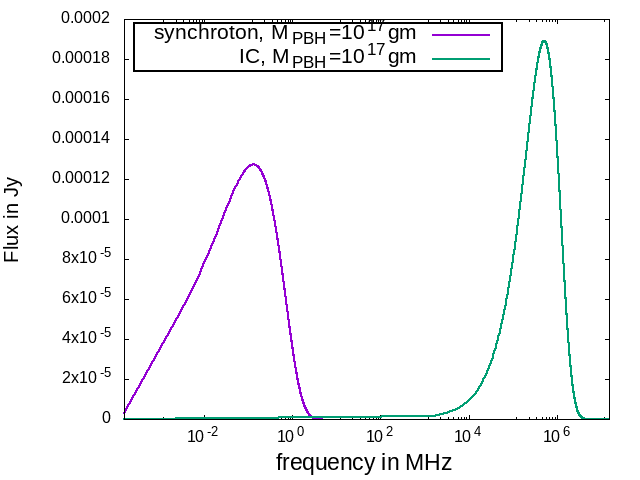}
\includegraphics[width=8cm,height=5cm]{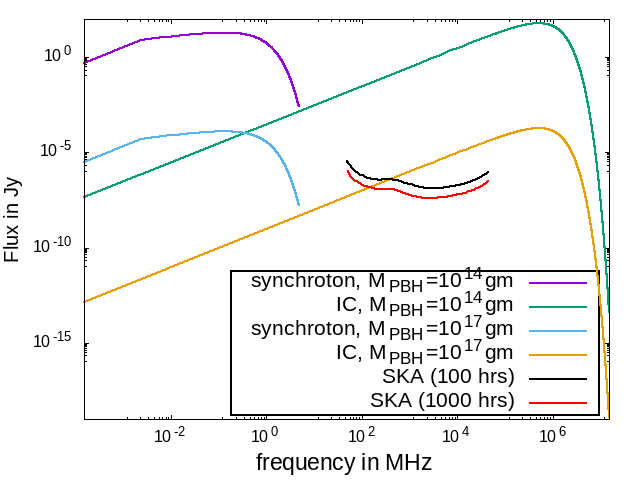}

\caption{{\it Left Panel:} Variation of the flux of synchrotron radiation (the purple line in the plot) and of IC radiation (the green line in the plot) with frequency $\nu$, generated due to the evaporation of PBH with mass $M_{\rm PBH}$=$10^{17}$ g. {\it Right Panel:} Variation of the flux for synchrotron radiation and Inverse Compton with frequency $\nu$, originated due to the evaporation of PBH with masses $M_{\rm PBH}$=$10^{17}$ g and 4$\times$ $10^{14}$ g. Sensitivity of SKA for 100 hrs and 1000 hrs are shown by the black and red curves respectively.
} 
\label{set2:mono}
\end{figure}

\begin{figure}
\centering
\includegraphics[width=8cm,height=5cm]{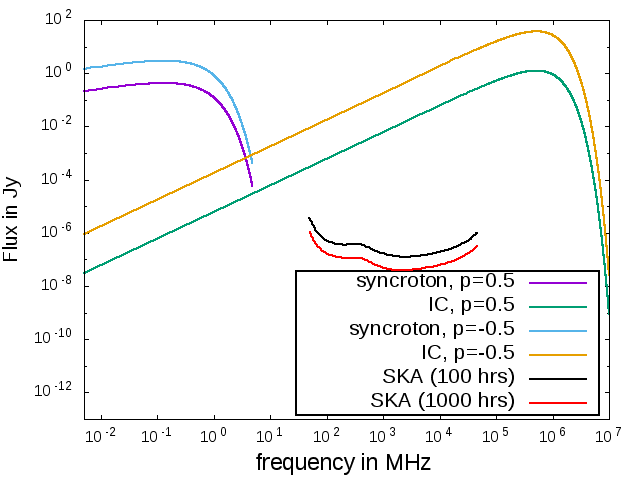}
\includegraphics[width=8cm,height=5cm]{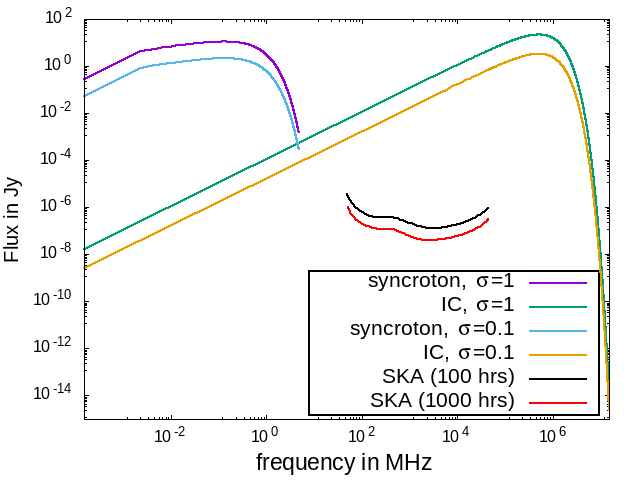}

\caption{{\it Left Panel:} Variation of the flux for synchrotron radiation and for IC radiation with frequency $\nu$, originated due to the evaporation of PBH with power law distribution of PBH mass where $M_{\rm min}$=4$\times 10^{14}$ g, $M_{\rm max}$=$10^{17}$ g and $p=-0.5, 0.5$. Sensitivity of SKA for 100 hrs and 1000 hrs are shown by the black and red curves respectively. {\it Right Panel:} Variation of the flux for synchrotron radiation and for IC radiation with frequency $\nu$, originated due to the evaporation of PBH with lognormal distribution of PBH mass where average of the mass distribution $\mu=5 \times 10^{15}$gm and variance $\sigma=0.1$ and $1$. Sensitivity of SKA for 100 hrs and 1000 hrs are shown by the black and red curves respectively.
}  
\label{set2: powerlog}
\end{figure}

\begin{figure}
\centering
\includegraphics[width=8cm,height=5cm]{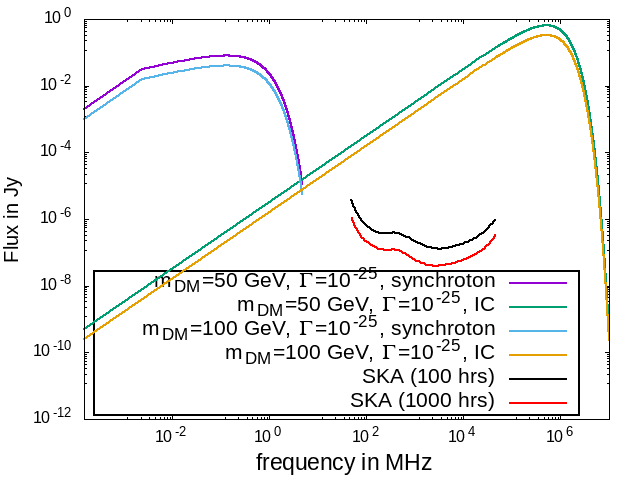}
\includegraphics[width=8cm,height=5cm]{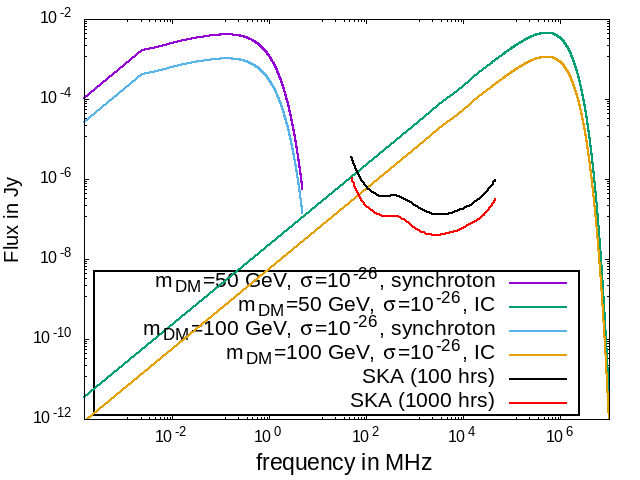}

\caption{{\it Left Panel:} Variation of the flux for synchrotron radiation and IC radiation with frequency $\nu$, originated due to the decay of DM with mass $m_{\rm DM} = 50$ GeV, 100 GeV and decay rate $\Gamma = 10^{-25}$s$^{-1}$. Sensitivity of SKA for 100 hrs and 1000 hrs are shown by the black and red curves respectively. {\it Right Panel:} Variation of the flux for synchrotron radiation and IC scattering with frequency $\nu$, originated due to the annihilation of DM with mass $m_{\rm DM} = 50$ GeV and 100 GeV and annihilation rate $\left\langle   \sigma v \right\rangle = 10^{-26} {\rm cm}^3$s$^{-1}$. Sensitivity of SKA for 100hrs and 1000hrs are shown by the black and red curves respectively.
} 
\label{set2: DM}
\end{figure}

\begin{figure}
\centering
\includegraphics[width=8cm,height=5cm]{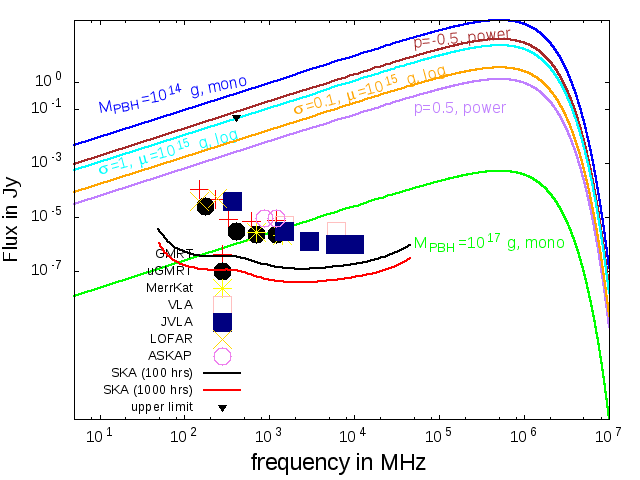}
\includegraphics[width=8cm,height=5cm]{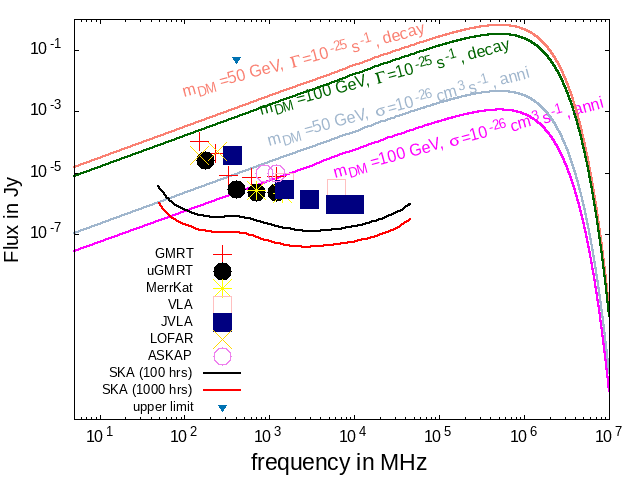}
\caption{{\it Left panel:} Variation of IC flux with frequency generated from the evaporation of PBH with different mass values and different mass distributions. {\it Right panel:} Variation of IC flux with frequency generated from the decay of DM and annihilation of DM with mass 50 GeV and 100 GeV. The fluxes are compared with the sensitivity of different radio frequency experiments i.e., GMRT, uGMRT, MeerKat, VLA, JVLA, LOFAR, ASAP, SKA (100 hrs), SKA (1000 hrs).}
\label{set2: exp}
\end{figure}

So far, we are assumed the $e^-/e^+$ produced as a result of PBH evaporation and DM decay or annihilation to be particles with energy in GeV order (high enrgy $e^-/e^+$ case). In this section, we focus on the synchrotron or IC radiation from such $e^-/e^+$ if their energies are in the order of MeV. To this end low energy electrons/positrons with energy $E\sim 2$ MeV are considered to compute as before, the synchrotron peak frequency and IC peak frequency.
For all our calculations in this section, we have considered $f_{\rm PBH}=10^{-7}$ and NFW profile of DM Halo unless otherwise mentioned. In the present calculations, synchrotron emissions are found to yield possible signals with the peak frequency at $\sim 10^{-1}$ MHz while the IC radiations produce radio signals with the peak frequency at $\sim 10^{5}$ MHz. Therefore, for low energy $e^-/ e^+$ the frequencies of the synchrotron signals are very low and not in the detectable range of present or upcoming radio telescopes but IC radiation appears to produce detectable radio signals. Hence, in this section, for the case of $e^-/e^+$ with energy in MeV range we have compared the flux densities from IC process only with different experimental data. Likewise the previous sections, here too five processes are considered, the end product $e^-/e^+$ of which may induce radio flux densities after being undergone IC scattering. These five processes are i) decay of PBH with monochromatic mass distribution, ii) decay of PBH with power law distribution of mass, iii) decay of PBH with lognormal mass distribution iv) decay of DM, v) annihilation of DM. The flux are then computed for each of the five cases and the results are plotted in Figs. \ref{set2:mono} - \ref{set2: exp}.

In the left panel of Fig. \ref{set2:mono} variations of the flux for synchrotron radiation (the purple line in the plot) and IC radiation (the green line in the plot) with frequency $\nu$, generated due to the evaporation of PBH with mass $M_{\rm PBH}$=$10^{17}$ g are shown while in the right panel of Fig. \ref{set2:mono} variations of the flux for synchrotron radiation and IC radiation with frequency $\nu$, originated due to the evaporation of PBH with masses $M_{\rm PBH}$=$10^{17}$ g and 4$\times$ $10^{14}$ g are compared. Sensitivities of SKA for 100 hrs and 1000 hrs are also indicated in the figure for comparison. As in Fig. \ref{fig:mono}, Fig. \ref{set2:mono} also shows smaller flux densities for larger PBH mass $10^{17}$ g than that for PBH mass of 4$\times 10^{14}$ g as expected. But comparing Figs. \ref{fig:mono} and \ref{set2:mono} it can be noted that the peak frequencies for both the synchrotron flux density and IC flux density are different for the two cases (as already discussed in the previous paragraph). It can also be noted that flux densities for Fig. \ref{set2:mono} are somewhat smaller than the flux densities in Fig. \ref{fig:mono}.

The flux calculations (synchrotron and IC) related to the PBH evaporation for power law and lognormal distributions of PBH are shown in Fig. \ref{set2: powerlog}. In the left panel of Fig. \ref{set2: powerlog} variations of the flux for synchrotron radiation and IC radiation with frequency $\nu$, produced from the evaporation of PBH with power law distribution of PBH mass where $M_{\rm min}$=4$\times 10^{14}$ g, $M_{\rm max}$=$10^{17}$ g and $p=-0.5, 0.5$ are plotted and sensitivity of SKA for 100 hrs and 1000 hrs are also shown by the black and red curves respectively. It can be compared with Fig. \ref{fig:powerlog} (left panel). Note that in this case (Fig. \ref{set2: powerlog} (left panel)) also larger flux densities arise for the case $p=-0.5$ as expected. The right panel of Fig. \ref{set2: powerlog} shows variation of the flux for synchrotron radiation and IC radiation with frequency $\nu$, from the evaporation of PBH with lognormal distribution of PBH mass. The average of the mass distribution $\mu$ is taken to be $\mu=5 \times 10^{15}$ g and the computations are made for two values of variances namely, $\sigma=0.1$ and $1$. Sensitivity of SKA for 100 hrs and 1000 hrs are also shown by the black and red curves respectively. One observes from this figure that larger flux densities are obtained when $\sigma=1$ (as in the case of Fig. \ref{fig:powerlog} (right panel)). Comparing Fig. \ref{fig:powerlog} and Fig. \ref{set2:mono} it can again be noted that the peak frequencies for synchrotron and IC radiation fluxes are different for the two figures and for low energy $e^-/e^+$ scenario flux densities have smaller values.


The low energy $e^-/e^+$ that may produce due to decay or annihilation of DM will now be considered. The synchrotron and IC fluxes from such $e^-/e^+$ are computed and the results are plotted in Fig. \ref{set2: DM}. In the left panel of Fig. \ref{set2: DM}, variations of the flux for synchrotron radiation and IC with frequency $\nu$, originated due to the decay of DM with mass $m_{\rm DM} = 50$ GeV, 100 GeV and decay rate $\Gamma = 10^{-25}$s$^{-1}$ are shown while in the right panel of Fig. \ref{set2: DM} variations of the same with frequency $\nu$, originated due to the annihilation of DM with mass $m_{\rm DM} = 50$ GeV and 100 GeV and annihilation rate $\left\langle   \sigma v \right\rangle = 10^{-26} {\rm cm}^3$s$^{-1}$ are plotted. Sensitivity of SKA for 100 hrs and 1000 hrs are shown by the black and red curves respectively. For both DM decay and DM annihilation cases flux densities for $m_{\rm DM} = 50$ GeV are larger than the case when DM mass $m_{\rm DM} = 100$ GeV is considered. Here too, it can be stated by comparing Fig. \ref{fig:DM} and Fig. \ref{set2: DM} that the peak frequencies are different for both the sets and flux densities due to $e^-/e^+$ with energy in the order of MeV are smaller than that for high energy $e^-/e^+$ case ($\sim$ GeV).

As demonstrated in Fig. \ref{fig:exp} (comparison of synchrotron flux densities with experimental data) in Sect. \ref{sect3.1}, in Fig. \ref{set2: exp}, we have compared the IC flux densities with the results (expected results) of present and upcoming radio telescopes. It is to be mentioned again that this calculations are for low energy $e^+/e^-$ which we consider to be decay/annihilation product of DM. In the left panel of Fig. \ref{set2: exp}, flux densities due to the decay of PBHs with different masses and mass distributions are compared and in the right panel of Fig. \ref{set2: exp} flux densities produced from DM decay/annihilation for $m_{\rm DM} = 50$ GeV, 100 GeV are compared. 
Similar to what have been shown in Fig. \ref{fig:exp}, a point labeled ``upper limit'' has also been shown in Fig. \ref{set2: exp} to indicate the upper limit of the radio flux densities at the Galactic Centre at frequency 408 MHz (50 mJy). This can also be noted from Fig. \ref{set2: exp} that all the computed flux for DM decay/annihilation case lie below the ``upper limit'' (right panel of Fig. \ref{set2: exp}) while some of the computed fluxes related to PBH have not crossed the ``upper limit''. In contrast, in similar plots of synchrotron flux densities with high energy $e^+/e^-$ in Fig. \ref{fig:exp} (left and right panels), all the calculated fluxes lie above the ``upper limit''. This shows that the IC induced fluxes inspired by PBH evaporation and DM decay/annihilation are in general lower for low energy $e^+/e^-$.

Also by comparing Fig. \ref{fig:exp} and Fig. \ref{set2: exp} it can be stated that the flux densities for the low energy $e^-/ e^+$ case are smaller than the flux densities for the high energy $e^+/e^-$ and thus larger number of plots lie below the upper limit provided by Jodrell Bank data (the ``upper limit'' point in Fig. \ref{set2: exp}).
\subsubsection{Constraints on the Parameters from Experimental Limits for Low Energy Electrons and Positrons}\label{boundset2}
\begin{figure}[H]
\centering
\includegraphics[width=8cm,height=5cm]{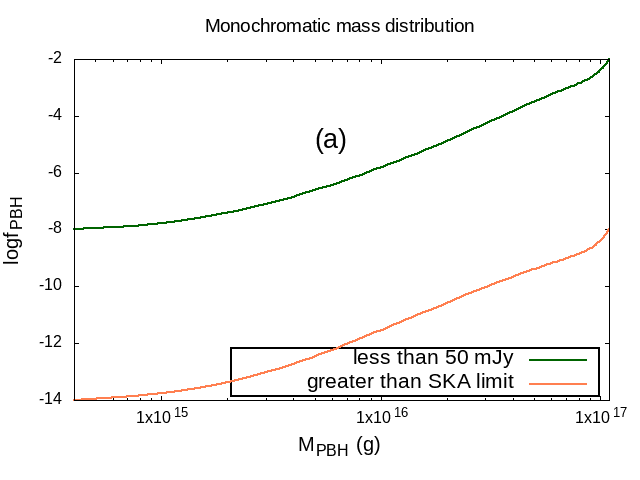}
\includegraphics[width=8cm,height=5cm]{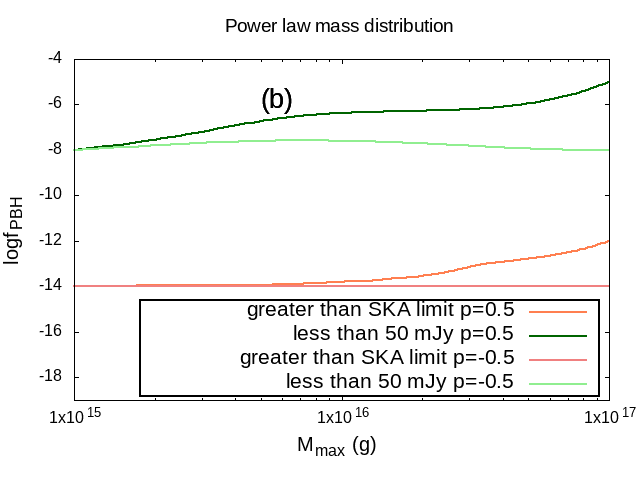}
\includegraphics[width=8cm,height=5cm]{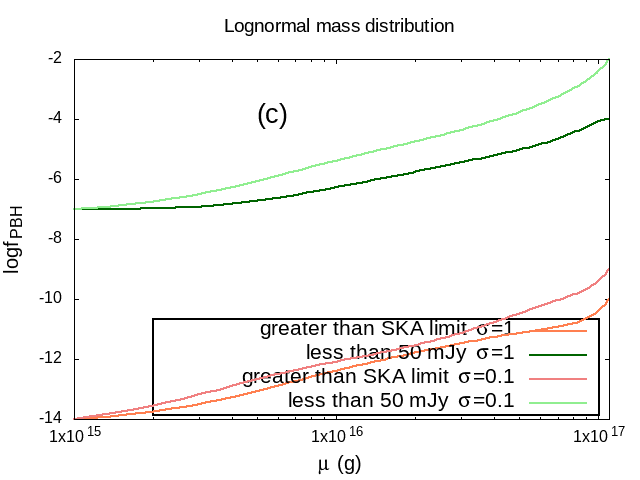}
\caption{(a) The upper limit and lower limit of $f_{\rm PBH}$ as a function of the PBH monochromatic mass. (b) The upper limit and lower limit of $f_{\rm PBH}$ as a function of the maximum mass for power law mass distribution of PBH. (c) The upper limit and lower limit of $f_{\rm PBH}$ as a function of the mean value of lognormal mass distribution of PBH.}
\label{limit_mono}
\end{figure}

\begin{figure}[H]
\centering
\includegraphics[width=7.5cm,height=5cm]{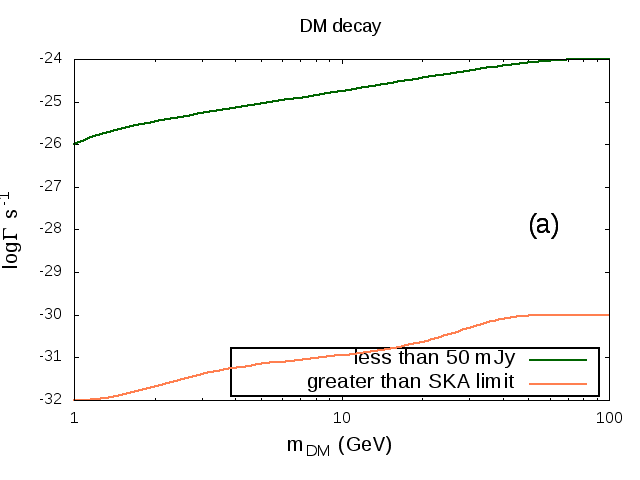}
\includegraphics[width=7.5cm,height=5cm]{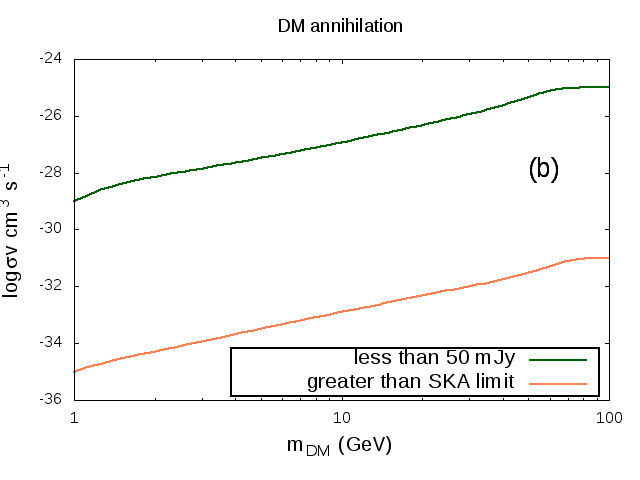}
\caption{(a) The upper limit and lower limit of $\Gamma$ as a function of the DM mass $m_{\rm DM}$ for the case of DM decay. (b) The upper limit and lower limit of  $\left\langle   \sigma v \right\rangle$ as a function of the DM mass $m_{\rm DM}$ for the case of DM annihilation.}
\label{limit_DM}
\end{figure}

In this section, we have constrained the parameters $f_{\rm PBH}$, $M_{\rm PBH}$, $\Gamma$,  $\left\langle   \sigma v \right\rangle$ and $m_{\rm DM}$ from the IC scattering of low energy $e^-/e^+$ emitted from the decay of PBH or decay/annihilation of DM. These constraints are obtained using the sensitivity of SKA and applying the condition that the experimental prediction for radio flux with 4 arcsec of the Galactic Centre can not be greater than 50 mJy at 408 MHz. The allowed limits of the parameters obtained using SKA sensitivity has been shown in Figs. \ref{limit_mono} and \ref{limit_DM} (the region above the orange lines) whereas the later condition is shown by the dark green lines in both Figs. \ref{limit_mono} and \ref{limit_DM}.
In Fig. \ref{limit_mono}(a), (b) and (c) the constraints are shown for the case of monochromatic mass distribution of PBH, power law mass function of PBH and lognormal mass function of PBH respectively. From Fig. \ref{limit_mono}(a) it can be observed that to produce detectable flux density in SKA, $f_{\rm PBH}$ can be as small as $10^{-14}$ and $10^{-9}$ respectively for PBH mass $4\times 10^{14}$ g and $10^{17}$ g but $f_{\rm PBH}$ should not be greater than $10^{-8}$ and $10^{-2}$ respectively for these PBH masses. In Fig. \ref{limit_mono}(b), we constrain the parameter space $f_{\rm PBH}$ with $M_{\rm  max}$ with the condition described earlier, for two power law index values namely $p=-0.5$ and $p=0.5$ while $M_{\rm min}$ is kept fixed at $M_{\rm min}=4\times 10^{14}$ g. In Fig. \ref{limit_mono}(c) the constraints on $f_{\rm PBH}$ are plotted for lognormal mass distribution of PBH with mean value of the mass $\mu$ for $\sigma=0.1, 1$ are shown. By comparing Fig. \ref{limit_mono1} with Fig. \ref{limit_mono} it can be noted that the allowed range of $f_{\rm PBH}$ would be modified when the emission of low energy $e^+,e^-$ with energy of the order of MeV are considered (low energy case) than the case of $e^+,e^-$ with high energy (order of GeV). Thus, the allowed range of PBH densities depend on the energy of produced $e^+,e^-$.

In Fig. \ref{limit_DM}(a) the constraints are shown for DM decay rate $\Gamma$ and DM mass $m_{\rm DM}$ for loe energy $e^-/e^+$ case. It is such that SKA would be able to detect the emitted radio flux for DM mass 1 GeV and 100 GeV even when decay rate is very small say $10^{-32}$ s$^{-1}$ or $10^{-30}$ s$^{-1}$ respectively. But 
\hspace{-5 mm}the upper limits of $\Gamma$ are $10^{-26}$ s$^{-1}$ and $10^{-24}$ s$^{-1}$ for mass values 1 GeV and 100 GeV. In Fig. \ref{limit_DM}(b) the constraints are shown for the parameter space of DM annihilation rate $\left\langle   \sigma v \right\rangle$ and DM mass $m_{\rm DM}$. It may be seen that SKA would be able to detect the emitted radio flux for DM mass 1 GeV and 100 GeV for annihilation rate $10^{-35}$ cm$^3$ s$^{-1}$ or $10^{-30}$ cm$^3$ s$^{-1}$ respectively. But the upper limits of $\left\langle   \sigma v \right\rangle$ are $10^{-29}$ cm$^3$ s$^{-1}$ and $10^{-25}$ cm$^3$ s$^{-1}$ for mass values 1 GeV and 100 GeV.

\section{Summary and Discussions} \label{discussion}
The primordial black hole or PBH that are believed to have been produced due to perturbation in very early Universe may evaporate via Hawking radiation and leave fermions, photons etc. The charged $e^-/e^+$ thus produced may emit synchrotron radiation while propagating through the Galactic magnetic field. These charge particles on the other hand may induce Inverse Compton (IC) scattering and emit radio signal. Similar situation may arise for the decay or self annihilation of DM producing $e^-/e^+$ as the end products. The synchrotron radiation or IC radiation can also be induced in such scenario. In this work these emissions are addressed and their detectabilities are explored. Variations of such signals with the distance from the Galactic Centre are also observed. Additionally, attempts have been made to constrain the parameters associated with the above processes using experimental limits or available data.
In this work we have considered five cases to compute such synchrotron radiations and IC radiations. These are i) decay of PBH with monochromatic mass distribution, ii) decay of PBH
with power law mass distribution, iii) decay of PBH with lognormal mass distribution iv) decay
of DM and v) annihilation of DM. A comparison has also been made for these five cases. Further, the detectability
of such fluxes in the present and upcoming radio wave experiments are discussed and bounds on the model
parameters are obtained. Constraints on the model parameters are calculated for two energy ranges of electrons, high energy ($\sim 0.1$ GeV) $e^-/e^+$ and low energy ($\sim 2$ MeV) $e^-/e^+$. Variations of the
fluxes with the distance from the Galactic Centre are also produced.

It is found that larger flux densities are obtained when contributions from smaller mass PBHs become more significant. This is also the case when DM with smaller masses decay or annihilate to produce such flux densities. It is noted that for high energy $e^-/ e^+$ emission, synchrotron radiations produce radio signals with the peak frequency
at $\sim 338$ MHz while the IC radiations give rise to ultraviolet to X-ray signals. Therefore, the calculated synchrotron flux densities for different masses of PBH, mass functions of PBH and also for different masses of DM are compared with data from upcoming and present radio telescopes like GMRT, uGMRT, MeerKat, VLA, JVLA, LOFAR, ASKAP, SKA (100 hrs), SKA (1000 hrs). It can also be predicted from the comparison that parameters used to compute the synchrotron radiations from the decay of PBH and decay/annihilation of DM would be tightly constrained as the upper limit of the radio flux densities at the Galactic Centre at frequency 408 MHz is 50 mJy.

Variations of the synchrotron flux with the distance from the Galactic Centre ($r$) are also calculated for different PBH masses, different PBH mass distributions and different DM masses. It is found that the synchrotron emission is highest at a particular distance and that peak distance depends on the PBH mass or DM mass or on the type of PBH distributions. For heavier PBHs (or DMs) one would obtain peak flux densities at smaller distances (or nearer to the Galactic Centre). Similarly, for power law mass distribution or lognormal mass distribution of PBH the peak flux densities are situated nearer to the Galactic Centre when contributions of the decay of lighter PBHs are negligible. Therefore, the measurement of the peak positions of such fluxes can be useful to determine the PBH mass or DM mass at the Galactic Centre region. Moreover, variations of the IC flux densities, originated due to the evaporation of PBHs with
monochromatic mass distribution, with the distance from the Galactic Centre are computed and it is found that variations of the synchrotron flux densities and IC flux densities with $r$ (the distance from the Galactic Centre) show different behaviour and hence fluxes produced from these two processes can be distinguishable. We have also shown such variations of synchrotron flux with $r$ for a different DM profile namely, EINASTO profile of DM and observed that the variations depend on the form of DM halo profile.

Additionally, limits on the model parameters are calculated from the SKA sensitivity
and from the Jodrell Bank’s experimental result (radio flux within 4 arcsec at the Galactic Centre should
not be greater than 50 mJy at 408 MHz). It can be observed from the present calculations that SKA would be able to detect the radio signals even if energy density of PBH $\Omega_{\rm PBH}$ is $\sim 10^{-14}$ times $\Omega_{\rm DM}$ for PBH mass distributions. But the upper limits on the $f_{\rm PBH}$ (PBH fraction) are tight, which means amount of PBHs is a small component of DM at the Centre of Galaxy. The upper limit and lower limit of decay rates ($\Gamma$) of DM as a function of the DM mass and annihilation rates of DM ($\left\langle   \sigma v \right\rangle$) as a function of the DM mass are also obtained. It is noted that SKA would be able to detect the signals even if $\Gamma$ lies between $10^{-32} - 10^{-30}$ s$^{-1}$ for mass range 1 - 100 GeV respectively but the upper limit of $\Gamma$ should be between $10^{-27} - 10^{-25}$ s$^{-1}$ for such mass range. Similarly, SKA will be able to detect the signals even if $\left\langle   \sigma v \right\rangle$ lies between $10^{-36} - 10^{-33}$ cm$^3$ s$^{-1}$ for mass range 1 - 100 GeV respectively but the upper limit of $\left\langle   \sigma v \right\rangle$ should be between $10^{-29}-10^{-27}$ cm$^3$ s$^{-1}$ for such mass range.

The synchrotron radiation and IC radiation arises due to the low energy $e^-, e^+$ originated from the above mentioned five scenarios are also computed in this work. In
this case synchrotron emissions produce signals with the peak frequency at $\sim 10^{-1}$ MHz while
the IC radiations produce radio signals with the peak frequency at $\sim 10^5$ MHz. Therefore, for low energy $e^-, e^+$,
we have compared the IC radiation flux densities with radio telescopes data. It is found that for  $e^-, e^+$ with low energy, the flux densities are smaller than the flux densities originated from high energy  $e^-, e^+$. Therefore, the constraints on the upper limit of $f_{\rm PBH}$, $\Gamma$ or $\left\langle   \sigma v \right\rangle$ are comparatively less severe. For lognormal mass distribution and mochromatic mass distribution of PBH, $f_{\rm PBH}$, fraction of PBH in the Centre of Galaxy, can be upto $10^{-2}$ and for power law mass distribution it is upto $10^{-4}$. The decay rate of DM can take values between $10^{-26} - 10^{-24}$ s$^{-1}$ and annihilation rate can lie between $10^{-29} - 10^{-25}$ cm$^3$ s$^{-1}$ for DM mass of 1 - 100 GeV.

In summary, in this work we have observed synchrotron flux densities and IC flux densities originated from the five above mentioned process at the Galactic Centre region. Fluxes computed for each process are compared. Further, the fluxes are compared with present and upcoming radio telescopes data and bounds on the model parameters are obtained from experimental predictions. The cases are compared for the two energy ranges of produced $e^-/e^+$, high energy (order of GeV) and low energy (order of MeV). It is also found that variations of synchrotron fluxes with the distance $r$ can be a useful probe to determine the mass of the PBH (or parameters of PBH mass functions) and mass of the DM. 

It can be mentioned here that observations in the Galactic Centre region in radio wavelengths are complicated due to the presence of other astrophysical sources besides PBHs and DM. One of the most significant radio sources of the GC is Sagittarius A$^*$ (Sgr A$^*$).
The measured integrated flux densities of Sgr A$^*$ by the VLA radio telescope are 0.71 Jy to 1.53 Jy for the frequencies 1.5 GHz to 41 GHz respectively \cite{JVLA}. On the other hand, the GMRT has observed the central region of the Milky Way at lower frequencies namely 580 MHz, 620 MHz and 1010 MHz for the detection of emission from Sgr A$^*$ \cite{Roy:2004th}. From the latter observations the flux density at 620 MHz is estimated to be 0.5$\pm$0.1 Jy while at 1010 MHz the same is estimated to be 0.6$\pm$0.12 Jy. Flux density at 580 MHz matches with that at 620 MHz within the error bar.
In our work, we have calculated the synchrotron and IC flux densities for the case when $e^-/e^+$ are produced due to the evaporation of PBH or as the possible end products of annihilation or decay of DM at the central region of our Galaxy. For high energy $e^-/e^+$ ($E\sim 0.1$ GeV), it is found in our calculation that the synchrotron emission gives radio signals with peak frequency at $\sim$ 338 MHz and the IC emission gives UV to X-ray emission while for low energy $e^-/e^+$ (E$\sim$ 2 MeV), the peak frequencies of synchrotron and IC emissions are $\sim 10 ^{-1}$ MHz and $\sim 10^5$ MHz respectively. Here, the calculated flux densities primarily depend on the PBH (or DM) mass and mass distributions. For PBH mass distributions (monochromatic, power law and log normal) the flux densities are found to vary from $O (10^{-5})$ Jy to $O(1)$ Jy and for the case of DM annihilation/decay the variations for the same are obtained from $O (10^{-3})$ Jy to $O(1)$ Jy. Thus, while observational results indicate particular flux densities of Sgr A$^*$ at particular frequencies, in our calculations a range of flux densities are obtained at a particular frequency depending on the mass values and mass distributions of PBH (or DM). Therefore, improvement in measuring the flux densities in the future observations could be useful to discriminate different radio sources at the GC. Moreover, in Ref. \cite{Diesing:2017,Capellupo:2017qwu} the variabilities of the emission of Sgr A$^*$ are observed with both time and frequency and it is found that Sgr A$^*$ flux density is highly variable with time (on an hourly time scale). But the flux densities originated either from PBH evaporation or from DM annihilation/decay would remain unchanged for such small time periods and thus would show constant flux densities with time in contrast to Sgr A$^*$.

\vskip 1cm
\noindent{\large \bf Acknowledgements}  

\noindent The authors would like to thank K. K. Datta for useful initial suggestions. One of the authors (U.M.) receives her fellowship grant from Council of Scientific \& Industrial Research (CSIR), Government of India as Senior Research Fellow (SRF) with the fellowship Grant No. 09/489(0106)/2017-EMR-I.

\normalem
\bibliography{reference}
\end{document}